\documentclass{emulateapj}
\usepackage{amssymb}
\usepackage{lscape}
\usepackage{longtable}
\begin{document}
\newcommand{\ha}{H$\alpha$}
\newcommand{\Ha}{\,{\rm H\alpha}}
\newcommand{\lya}{Ly$\alpha$}
\newcommand{\lyb}{Ly$\beta$}
\newcommand{\za}{$z_{\rm abs}$}
\newcommand{\ze}{$z_{\rm em}$}
\newcommand{\cmtwo}{cm$^{-2}$}
\newcommand{\nhi}{$N$(H$^0$)}
\newcommand{\degpoint}{\mbox{$^\circ\mskip-7.0mu.\,$}}
\newcommand{\kms}{\,km~s$^{-1}$}      
\newcommand{\minpoint}{\mbox{$'\mskip-4.7mu.\mskip0.8mu$}}
\newcommand{\peryr}{\mbox{$\>\rm yr^{-1}$}}
\newcommand{\secpoint}{\mbox{$''\mskip-7.6mu.\,$}}
\newcommand{\sqdeg}{\mbox{${\rm deg}^2$}}
\newcommand{\squig}{\sim\!\!}
\newcommand{\subsun}{\mbox{$_{\odot}$}}
\newcommand{\et}{{\rm et al.}~}
\newcommand{\msun}{\,{\rm M_\odot}}
\def\ltsima{$\; \buildrel < \over \sim \;$}
\def\simlt{\lower.5ex\hbox{\ltsima}}
\def\gtsima{$\; \buildrel > \over \sim \;$}
\def\simgt{\lower.5ex\hbox{\gtsima}}
\def\arcs{$''~$}
\def\arcm{$'~$}
\def\erf{\mathop{\rm erf}}
\def\erfc{\mathop{\rm erfc}}

\title{Chemical Abundances of DEEP2 Star-forming Galaxies at $1.0 < z < 1.5$\altaffilmark{1}}

\slugcomment{DRAFT: \today}
\author{\sc Alice E. Shapley\altaffilmark{2}, Alison L. Coil, \& Chung-Pei Ma}
\affil{University of California, Berkeley, Department of Astronomy, 601 Campbell Hall,
Berkeley, CA 94720}
\author{\sc Kevin Bundy}
\affil{California Institute of Technology, MS 105-24, Pasadena, CA 91125}

\altaffiltext{1}{Based, in part, on data obtained at the 
W.M. Keck Observatory, which 
is operated as a scientific partnership among the California Institute of 
Technology, the
University of California, and NASA, and was made possible by the generous 
financial
support of the W.M. Keck Foundation.} 
\altaffiltext{2}{Miller Fellow}

\begin{abstract}
We present the results of near-infrared spectroscopic observations for a
sample of 12 star-forming galaxies at $1.0 < z < 1.5$, drawn from the
DEEP2 Galaxy Redshift Survey.  H$\beta$, [OIII], H$\alpha$, and [NII]
emission-line fluxes are measured for these galaxies. Application of the
$O3N2$ and $N2$ strong-line abundance indicators implies average gas-phase
oxygen abundances of $50-80$\% solar.  We find preliminary evidence of
luminosity-metallicity $(L-Z)$ and mass-metallicity $(M-Z)$ relationships
within our sample, which spans from $M_B=-20.3$ to $-23.1$ in rest-frame
optical luminosity, and from $4\times 10^{9}$ to $2\times
10^{11}M_{\odot}$ in stellar mass. At fixed oxygen abundance, these
relationships are displaced from the local ones by several magnitudes
towards brighter absolute $B$-band luminosity and more than an order of
magnitude towards larger stellar mass. If individual DEEP2 galaxies in our
sample follow the observed global evolution in the $B$-band luminosity
function of blue galaxies between $z\sim 1$ and $z\sim 0$
\citep{willmer2005}, they will fade on average by $\sim 1.3$ magnitudes in
$M_B$. To fall on local $(L-Z)$ and $(M-Z)$ relationships, these galaxies
must increase by a factor of $6-7$ in $M/L_B$ between $z\sim 1$ and $z\sim
0$, and by factor of two in both stellar mass and metallicity. Such
concurrent increases in stellar mass and metallicity are consistent with
the expectations of a ``closed-box'' chemical evolution model, in which
the effects of feedback and large-scale outflows are not important. While
$K_s<20.0$ $z\sim 2$ star-forming galaxies have similar [NII]/Ha ratios
and rest-frame optical luminosities to those of the DEEP2 galaxies
presented here, their higher $M/L_B$ ratios and clustering strengths
indicate that they will experience different evolutionary paths to $z\sim
0$. Finally, emission line diagnostic ratios indicate that the $z> 1$
DEEP2 galaxies in our sample are significantly offset from the excitation
sequence observed in nearby H~II regions and SDSS emission-line galaxies.
This offset implies that physical conditions are different in the H~II
regions of distant galaxies hosting intense star formation, and may affect
the chemical abundances derived from strong-line ratios for such objects.

\end{abstract}

\section{Introduction}
\label{sec:intro}

While the $\Lambda$CDM framework provides detailed predictions for the
formation of dark matter structure, deep questions remain about the
baryonic physics of galaxy formation. Using statistical samples of
galaxies, it is now possible to apply critical observational tests to
numerical simulations and semi-analytic models of galaxy formation and
evolution over almost 90\% of the age of the universe. Important tests
include the determination of the global star-formation-rate and
stellar-mass density as a function of redshift
\citep{schiminovich2005,steidel1999,giavalisco2004,bunker2004,dickinson2003,fontana2004,drory2005}.
Evolution in the abundance of heavy elements in galaxies and the
intergalactic medium (IGM) provides an independent probe of the
star-formation history of the universe \citep{pei1998}, as the degree of
metal enrichment reflects the integrated products of star-formation. The
relationships among chemical enrichment, luminosity, and mass in galaxies
as a function of redshift highlight the importance of star-formation
feedback \citep{tremonti2004,kobulnicky2003}. Furthermore, the degree of
chemical enrichment in distant star-forming galaxies grants insight into
how populations of high-redshift galaxies map onto samples in the local
universe \citep{shapley2004}. Chemical abundances therefore serve as a
fundamental metric of the galaxy formation process.

The rest-frame optical integrated spectra of star-forming galaxies contain
several strong emission lines, including [OII]$\lambda\lambda 3726,3729$,
H$\beta$, [OIII]$\lambda\lambda 5007,4959$, H$\alpha$,
[NII]$\lambda\lambda 6548,6584$, and [SII]$\lambda\lambda 6717,6731$. The
relative strengths of these features indicate the physical conditions in
the H~II regions contained in the galaxies. One of the most important
quantities derived from the relative emission line strengths is the
average H~II-region metallicity. Both detailed photo-ionization models and
empirical calibrations have been used to relate the ratios of strong
emission lines to chemical abundances
\citep{pagel1979,evans1985,kewley2002,pp2004}. Locally, the metallicities
of star-forming galaxies have recently been measured by
\citet{tremonti2004}, who use a sample of $>53,000$ emission-line spectra
of galaxies selected from the Sloan Digital Sky Survey (SDSS) to
investigate the luminosity-metallicity and mass-metallicity relationships.
This sample is two orders of magnitude larger than previously existing
ones, and serves as a robust local comparison point for evolutionary
studies. At intermediate redshifts ($z<1$), there are several different
samples of star-forming galaxies with chemical abundances measurements,
the largest of which is presented in \citet{kobulnicky2004}, with $\sim
200$ galaxies at $0.30 \leq z < 0.94$.  Near-infrared spectroscopic
observations of rest-frame optical emission lines have been obtained for
galaxies drawn from several recently-discovered populations at $z\geq2$.
There are now rough chemical abundance measurements for $\sim 30$ objects
selected by their rest-frame UV luminosities and colors
\citep{pettini2001,shapley2004}, submillimeter emission
\citep{swinbank2004,tecza2004}, or observed $J-K$ colors
\citep{vandokkum2004,vandokkum2005}. However, there is still a glaring gap
in our knowledge of the metallicities of star-forming galaxies in the
redshift range $1.0 \leq z \leq 2.0$, an exciting era that hosts the
emergence of the Hubble sequence of disk and elliptical galaxies and the
build-up of most of the stellar mass in the universe
\citep{dickinson2003}.

Here we present near-infrared spectroscopic observations for a pilot
sample of 12 star-forming galaxies in this critical redshift range. The
galaxies are drawn from the DEEP2 Galaxy Redshift Survey
\citep{davis2003}, which contains $\sim 10000$ galaxies at $1.0\leq z \leq
1.5$, an unprecedented spectroscopic sample for addressing this large gap
in our knowledge of galaxy evolution. At these redshifts, the H~II region
lines of interest are shifted to $0.97-1.65 \mu$m, where bright sky lines
and regions of strong atmospheric absorption cause much of the spectral
real-estate to be unavailable. Therefore, the precise [OII] redshifts
provided by the DEEP2 survey are crucial for selecting objects whose
H$\beta$, [OIII], H$\alpha$, and [NII] lines will fall at optimum
wavelengths with respect to sky lines and atmospheric absorption. We have
found that two relatively narrow redshift windows within the DEEP2 survey
are ideal in this respect: $0.96 \leq z \leq 1.05$ and $1.36 \leq z \leq
1.50$.  We measure H$\alpha$ and [NII]$\lambda 6584$ 
fluxes for the entire pilot sample,
and additional [OIII]$\lambda 5007$ and H$\beta$ fluxes 
for a subsample of five. Even
with a small sample, we discern significant evolution between $z\sim 0$
and $z\geq 1.0$ in the relationships among chemical abundance, luminosity
and mass, as well as evidence for striking differences in the H~II region
physical conditions in distant star-forming galaxies. These results
highlight the promise of future $z\geq 1$ chemical abundance studies using
near-infrared spectrographs with wider simultaneous wavelength baselines
and multi-object capabilities.

We describe the DEEP2 target sample, and near-IR spectroscopic
observations and measurements in \S\ref{sec:obs}.  In \S\ref{sec:oh}, we
present the oxygen abundances derived from measurements of [OIII],
H$\beta$, H$\alpha$, and [NII] emission lines. Evolution in the
luminosity-metallicity and mass-metallicity relationships is discussed in
\S\ref{sec:lzmz}, while the evidence for differences in $z\sim 1$ H~II
region physical conditions is presented in \S\ref{sec:bpt}. Finally, in
\S\ref{sec:summary}, we summarize our principal conclusions. A cosmology
with $\Omega_m=0.3$, $\Omega_{\Lambda}=0.7$, and $h=0.7$ is assumed
throughout.

\section{Sample \& Observations}
\label{sec:obs}

\subsection{DEEP2 Target Sample}
\label{sec:obsdeep}

\begin{figure*}
\plotone{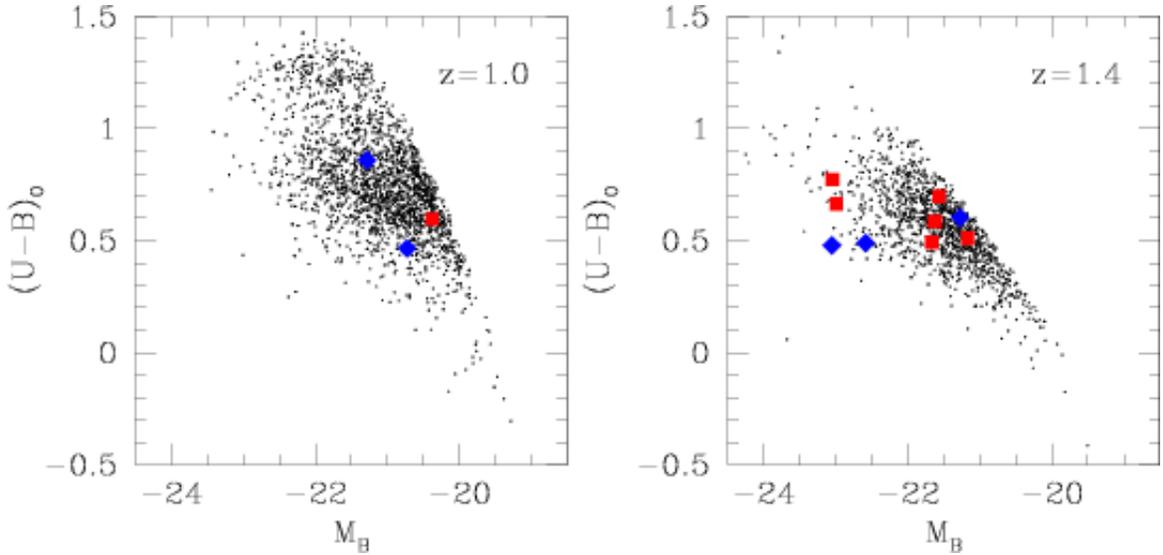}
\caption{DEEP2 Color-Magnitude Diagrams. These rest-frame $(U-B)_0$ vs.
$M_B$ color magnitude diagrams are for DEEP2 galaxies at $0.96\leq z\leq
1.05$ (left) and $1.36\leq z \leq 1.50$ (right). The target galaxies with
NIRSPEC spectra are indicated with larger symbols in each plot. Those with
the full set of H$\beta$, [OIII], H$\alpha$ and [NII] are indicated with
blue diamonds, while objects with only [NII] and H$\alpha$ measurements
are indicated with red squares. The set of $z\sim 1.4$ NIRSPEC targets
spans almost the full range in $M_B$ contained in the DEEP2 sample at that
redshift, while the $z\sim 1.0$ targets probe only the faint edge of the
range in the DEEP2 sample at $z\sim 1.0$.  In future observations, we will
attempt to span a larger range in $M_B$ at $z\sim 1.0$. A color bimodality
is found in the DEEP2 sample. All NIRSPEC targets were drawn from the blue
``cloud'' of the color bimodality.
}
\label{fig:cmdiagram}
\end{figure*}

The sample presented here was drawn from the DEEP2 Galaxy Redshift Survey
\citep[hereafter DEEP2;][]{davis2003}. DEEP2 is a three-year project using
80 nights on the DEIMOS spectrograph \citep{faber2003} on the Keck~II
telescope to survey optical galaxies at $z\sim 1$.  The DEEP2 survey was
designed to study both large-scale structure and galaxy evolution in a
large cosmic volume and is producing a dense mapping of the spatial
locations of $\sim$40,000 galaxies brighter than $R_{AB}=24.1$ with
redshifts $0.7<z<1.5$.  Photometric data were taken in $B, R$ and
$I$-bands with the CFH12k camera on the 3.6-m Canada-France-Hawaii
telescope \citep[see][for details]{coil2004b}. Follow-up near-IR
$K_s$-band photometry has been obtained in portions of all four DEEP2
survey fields. More than 15000 DEEP2 galaxies with spectroscopic redshifts
have measured $K_s$ magnitudes, which, when combined with optical data,
probe stellar masses and stellar population parameters (Bundy et al., in
preparation).  The DEEP2 spectroscopic dataset consists of DEIMOS spectra
typically covering the spectral range $6400 \leq \lambda \leq 9100$~\AA.
The overall success rate in measuring precise spectrosopic redshifts is
$\sim 80$\%. The [OII]~$\lambda\lambda 3726,3729$ emission doublet is used
to find redshifts over the entire DEEP2 range for more than $\sim 85$\% of
the galaxies in the sample with successful spectroscopic identifications.
The remainder of the spectroscopic sample have redshifts identified based
on stellar absorption lines alone.

To estimate accurate chemical abundances and H~II region physical
conditions, observations of several strong H~II region emission lines are
required -- ideally at least [OII], H$\beta$, [OIII], H$\alpha$, and
[NII]. At $z\geq 0.85$, however, the only strong H~II region emission
feature contained in the DEIMOS window is the [OII] doublet.  To measure
longer-wavelength emission lines at $z\geq 1$, we therefore require
near-IR spectroscopy. Because of the effects of strong atmospheric
absorption in the near-IR, the full set of H$\beta$, [OIII], H$\alpha$,
and [NII] is not available for all redshifts between $z=1.0$ and $1.5$. We
therefore targeted two differential redshift windows within the larger
DEEP2 redshift distribution:  $0.96 \leq z \leq 1.05$ and $1.36 \leq z
\leq 1.50$. Within these redshift windows, it is possible to measure all
of the emission lines of interest in the near-IR.  Furthermore, we
preferentially observed galaxies with the maximum number of emission lines
free from contamination by bright sky OH lines, based on
previously-determined [OII] redshifts. The pilot sample presented here
contains 3 galaxies at $z\sim 1.0$ and 9 galaxies at $z\sim 1.4$. The
disparity in size of the two subsamples was a result of limited observing
time, and will be rectified with future observations.

Galaxies were selected to span the full range of absolute $B$-luminosities
in the DEEP2 survey, from $M_B\sim -20$ to $-23$, in order to probe the
metallicity-luminosity relationship over an interesting dynamic range.  
As shown in Figure~\ref{fig:cmdiagram}, the larger set of $z\sim 1.4$
galaxies does span the full range of $M_B$ in the DEEP2 sample available
at that redshift, while the smaller set of $z\sim 1$ targets happens to be
drawn from the faint end of the luminosity function.  All targets have
restframe $(U-B)_0$ colors blueward of the minimum in the observed color
bi-modality in the DEEP2 survey. Finally, in order to maximize the
longslit observing efficiency, we mostly targeted pairs of galaxies in
each redshift window, with separations of up to 25".

\subsection{Near-IR Spectroscopy}
\label{sec:obsnearir}

The Keck~II/NIRSPEC observations were conducted on 5 and 24 October 2004.
At $z\sim 1.0-1.5$, the full complement of H$\beta$, [OIII], H$\alpha$,
and [NII] cannot be observed simultaneously -- two filter setups are
required to cover all the lines. For objects at $z\sim 1.4$, we used the
NIRSPEC~5 filter (similar to $H$ band) to observe H$\alpha$ and [NII], and
the NIRSPEC~2 filter (similar to a blue $J$ band) to observe [OIII] and
H$\beta$. For objects at $z\sim 1.0$, we used the NIRSPEC~4 filter
(similar to a red $J$ band) for observations of H$\alpha$ and [NII], and
the NIRSPEC~1 filter ($\Delta\lambda = 0.95 - 1.10 \mu\mbox{m}$)  for
observations of [OIII] and H$\beta$. For all observations, we used a
0\secpoint76~$\times$~42'' long-slit. The resulting spectral resolution
determined from sky lines is $\sim 10.5$~\AA\ in the NIRSPEC~5 and
NIRSPEC~4 filters (relevant for H$\alpha$ and [NII] measurements) and
$\sim 8$~\AA\ in the NIRSPEC~2 and NIRSPEC~1 filters (relevant for [OIII]
and H$\beta$ measurements). Blind offsets from nearby bright stars were
used to acquire our target objects, which are too faint to acquire
directly onto the slit. In most cases, the slit position angle was
determined by the attempt to fit two galaxies on the slit. On 5 October,
in the face of intermittent failures of the NIRSPEC field rotator, whose
proper functioning was required to target pairs, we could only acquire
single objects in the following cases: 32020728, 42008219, 42059947.
Targets were typically observed for 2 or 3~$\times$~900 seconds in each
filter. \footnote{Exceptions are 42008627, which was only detected in a
single 900 second exposure, out of three attempts, due to confirmed image
rotator problems; and 32025514, for which data were combined from two
separate nights, totaling $6 \times 900$~seconds. 32025514 was reobserved
in order to detect its pair companion, 32100778, which was missed on the
first attempt, due to rotator problems.}
 
Photometric conditions and seeing were variable throughout both nights,
with seeing ranging from 0\secpoint5 to 1\secpoint0 in the near-IR. We
targeted a total of 12 DEEP2 galaxies, successfully measuring 
[NII]$\lambda 6584$ and
H$\alpha$ for the entire sample, and [OIII] $\lambda 5007$ and H$\beta$
for 5 out of 12. For the remaining seven objects without measured
[OIII]~5007/H$\beta$ ratios, observations of those lines were either not
attempted because of time constraints, or affected by significantly (and
variable)  non-photometric conditions. A summary of the observations
including target coordinates, redshifts, optical and near-IR photometry,
and total exposure times is given in Table~1. The
two-dimensional galaxy spectral images were then reduced, extracted to one
dimension along with 1-$\sigma$ error spectra, and flux-calibrated with
observations of A stars, according to the procedures outlined in detail in
\citet{erb2003}.

\subsection{Measurements}
\label{sec:obsmeas}

\begin{figure*}
\epsscale{1.2}
\plotone{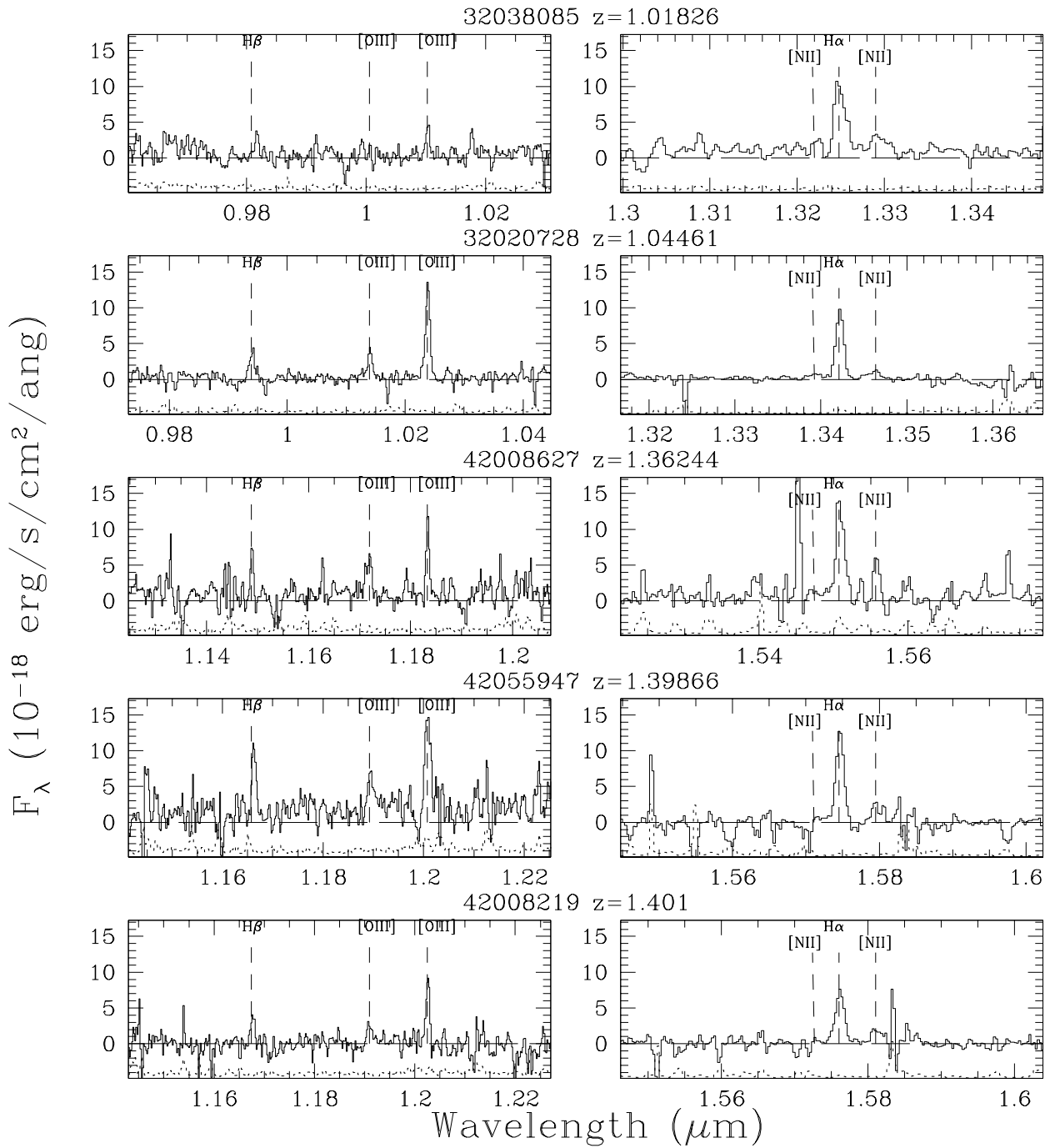}
\epsscale{1.0}
\caption{NIRSPEC spectra of DEEP2 galaxies at $1.0 \leq z < 1.5$.
Spectra for these five objects cover all of H$\beta$, [OIII], H$\alpha$
and [NII], enabling measurements of the $O3N2$ abundance indicator and the
emission-line diagnostic of [OIII]/H$\beta$ vs. [NII]/H$\alpha$. For the
objects at $z\sim 1.0$ (top two panels), H$\beta$ and [OIII] are observed
in the NIRSPEC~1 filter, while H$\alpha$ and [NII] are observed in the
NIRSPEC~4 filter (equivalent to a red $J$-band). For objects at $z\sim
1.4$ (bottom three panels), H$\beta$ and [OIII] are observed in the
NIRSPEC~2 filter (equivalent to a blue $J$-band), while H$\alpha$ and
[NII] are observed in the NIRSPEC~5 filter (equivalent to $H$-band). In
each panel, the dotted line shows the $1\sigma$ error spectrum, offset
vertically by $-5\times10^{-18}\mbox{ erg s}^{-1}\mbox{cm}^{-2}\mbox{ang}^{-1}$ 
for clarity. The NIRSPEC~5 spectrum for the object,
42008627, consists of only a single exposure, and exhibits more
cosmetic residuals due to associated limitations in the cosmic-ray zapping.
The line flux measurements of interest should not be
adversely affected, however.
}
\label{fig:oiiihalphaspec}
\end{figure*}

\begin{figure*}
\epsscale{1.2}
\plotone{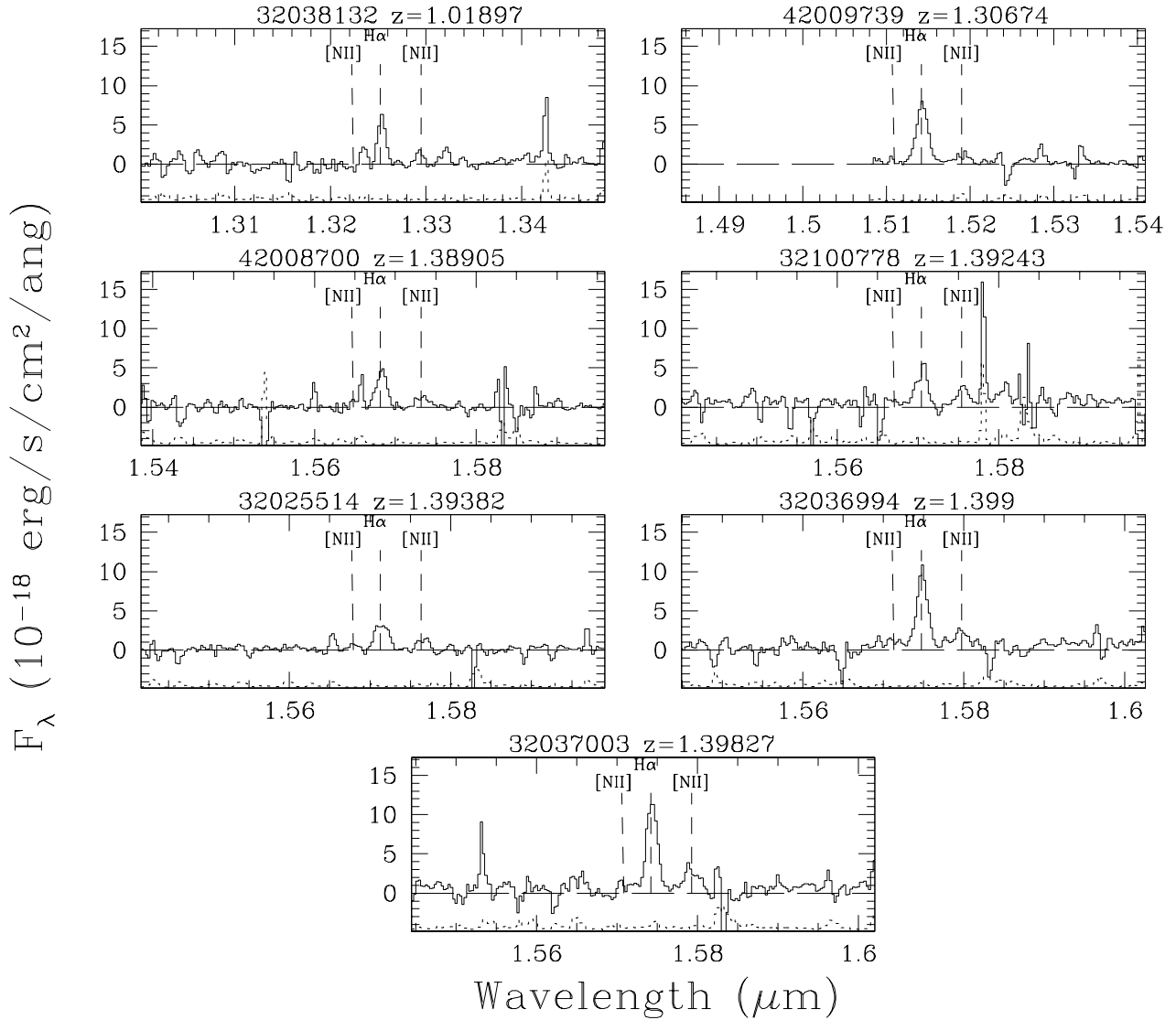}
\epsscale{1.0}
\caption{NIRSPEC spectra of DEEP2 galaxies at $1.0 \leq z < 1.5$.
Spectra for these seven objects only cover H$\alpha$ and [NII], which
enable crude measurements of chemical abundance with the $N2$ indicator.
H$\alpha$ and [NII] are observed in the NIRSPEC~4 filter (equivalent to a
red $J$-band) for the object at $z\sim 1.0$, and in the NIRSPEC~5 filter
(equivalent to $H$-band) for objects at $z\sim 1.4$. In each panel, the
dotted line shows the $1\sigma$ error spectrum, offset vertically by
$-5\times10^{-18}\mbox{ erg s}^{-1}\mbox{cm}^{-2}\mbox{ang}^{-1}$ for
clarity. The apparent spike in the spectrum of 32037003 at $1.553 \mu$m
is due to an insufficiently masked cosmic ray. Again,
the line flux measurements of interest should not be adversely affected.
}
\label{fig:onlyhalphaspec}
\end{figure*}

One-dimensional, flux-calibrated NIRSPEC spectra are shown in
Figures~\ref{fig:oiiihalphaspec} and \ref{fig:onlyhalphaspec}
for our entire sample. We show spectra for
objects with the full set of [OIII], H$\beta$, H$\alpha$, and [NII]
emission lines in two NIRSPEC filters, as well as for objects with
single-filter observations of [NII] and H$\alpha$ alone. $\Ha$ and 
[NII]$\lambda 6584$ emission line fluxes were determined by first 
fitting a one dimensional
Gaussian profile to the higher S/N $\Ha$ feature to obtain the redshift
and FWHM (in wavelength). The $\Ha$ redshift and FWHM were then used to
constrain the fit to the [NII] emission line. This procedure is based on
the assumption that the $\Ha$ and [NII] lines have exactly the same
redshift and FWHM, but that the $\Ha$ line offers a higher S/N estimate of
these parameters. The [NII]$\lambda 6548$ line was too faint to
measure for most of the objects in our sample.
[OIII]$\lambda 5007$ and H$\beta$ fluxes were
determined with independent fits to the line centroids and widths. For
most objects, the H$\alpha$, [OIII]$\lambda 5007$, 
and H$\beta$ redshifts agree to
within $\Delta z=0.0004$ ($\Delta v=50-60$~\kms\ at $z=1.0-1.4$).

\newpage

Though not the focus of the current work, we mention in passing that
10 out of 12 of the H$\alpha$ linewidths for our sample are resolved 
and range from $\sigma_v\sim 50$~\kms\ up to a
maximum of $\sigma_v\sim 125$~\kms. Two additional
galaxies have H$\alpha$ linewidths narrower or equal to
the instrumental resolution as estimated from the widths
of sky lines. The sample median is
$\sigma_{v,med}=80$~\kms. The median H$\alpha$ linewidth is consistent
with the median [OII] linewidth measured from the DEIMOS spectra of these
objects.  While slit position angles were chosen for the purpose of
targeting two objects simultaneously, regardless of the continuum or
[OII]-emission-line spatial extent or morphology, we find that three
objects (32037003, 32028085, and 42055947) have spatially-extended and
tilted emission-line profiles, indicative of rotation. Since we do not
have detailed morphological information for our sample, we do not further
interpret the tilted emission lines with a model for disk rotation.

The measured H$\alpha$ fluxes range from $5.6\times 10^{-17} \mbox{ erg
s}^{-1}\mbox{cm}^{-2}$ to $1.9 \times 10^{-16} \mbox{ erg
s}^{-1}\mbox{cm}^{-2}$. The mean flux for the sample at $z\sim 1.0$ is
$1.0 \times 10^{-16} \mbox{ erg s}^{-1}\mbox{cm}^{-2}$, corresponding to
an H$\alpha$ star-formation rate -- uncorrected for dust extinction -- of
$4 M_{\odot}\mbox{yr}^{-1}$, using the calibration of
\citet{kennicutt1998}. The mean for the sample at $z\sim 1.4$ is
$1.3\times 10^{-16} \mbox{ erg s}^{-1}\mbox{cm}^{-2}$, which corresponds
to an H$\alpha$ star-formation rate of $12 M_{\odot}\mbox{yr}^{-1}$. These
characteristic H$\alpha$ star-formation rates are discussed further in
section~\ref{sec:bptphys}. The close consistency between the flux
calibrations from standard stars taken on the two separate nights of
observations indicates that the standard stars were observed under
photometric conditions. The variability of conditions throughout the two
nights therefore translates into lower limits on the measured H$\alpha$
line fluxes and associated star-formation rates when the standard star
calibrations obtained during photometric conditions are applied to our
galaxy observations.  Even in photometric conditions, there are several
sources of systematic error affecting our absolute line flux measurements.  
These include imperfect centering of objects in the slit, variations in
the quality of seeing, and light lost because of the mismatch between the
slit width and the size of the DEEP2 targets, which are typically
0\secpoint5 in R-band (rest-frame near-UV) half-light radius.
\citet{erb2003} estimate that these sources of error amount to at least
$\sim 25$\% in absolute flux. For the remainder of the discussion, we
therefore emphasize the measured line {\it ratios}, [NII]~6584/H$\alpha$ and
[OIII]~5007/H$\beta$, which should be unaffected by uncertainties in
flux-calibration or the other systematics listed above. The uncertainty in
these line ratios is dominated by random error at the level of $\sim
10$\%.  Furthermore, since each line ratio is calculated from emission
lines very closely spaced in wavelength, its value is independent of the
magnitude of dust extinction. As shown in Table~2, the
measured [NII]~6584/H$\alpha$ ratios range from $0.10 - 0.44$, with an average
of $\langle \rm{[NII]~6584/H}\alpha \rangle = 0.25$. For the five galaxies with
measured [OIII]$\lambda 5007$ and H$\beta$ fluxes, 
the average [OIII]~5007/H$\beta$ ratio is
$\langle \rm{[OIII]~5007/H}\beta \rangle = 2.2$, ranging from $1.5-3.1$. The
average [NII]~6584/H$\alpha$ ratio for these five galaxies is 0.24, consistent
with the average for the total sample of 12.
Hereafter, we shall generally use ``[NII]/H$\alpha$'' to refer to the measured
ratio between [NII]$\lambda 6584$ and H$\alpha$, and 
``[OIII]/H$\beta$'' for the measured ratio between
[OIII]$\lambda 5007$ and H$\beta$.

\section{The Oxygen Abundance}
\label{sec:oh}

In distant star-forming galaxies, measuring the relative strengths of
strong emission lines is the only viable method for estimating the H~II
region gas-phase oxygen abundance \citep{kobulnicky1999,pettini2001}.  
The lines typically used include some subset of [OII]~$\lambda 3727$,
H$\beta$, [OIII]~$\lambda\lambda 4959, 5007$, H$\alpha$, and
[NII]~$\lambda 6584$. Empirical relations between strong-line ratios and
chemical abundance have been calibrated using local H~II regions with
measured electron temperatures, and therefore direct abundance
determinations \citep{pp2004}. Application of these relationships at
high-redshift rests on the assumption that the integrated spectra of
distant galaxies with high star-formation rates can be treated
equivalently to the spectra of individual local H~II regions used to
calibrate the strong-line abundance indicators. We will return to this
question in section~\ref{sec:bpt}.

\citet{pp2004} demonstrate that a reasonably accurate estimate ($\pm 0.14$
dex) of the H~II region oxygen abundance is possible with measurement of
the ``$O3N2$'' ratio $\equiv$ log(([OIII] 5007/H$\beta$)/([NII]
6584/H$\alpha$)). One major benefit of this abundance indicator over,
e.g., the indicator R$_{23} \equiv$~log(([OIII]+[OII])/H$\beta$), is that
$O3N2$ is relatively independent of flux calibration and uncertainties in
dust extinction, since it is derived from a pair of ratios of
closely-spaced emission lines. With all of the systematic uncertainties in
absolute line flux measurements described above, and including the
variable photometric conditions that characterized our observing runs, the
advantages of the $O3N2$ indicator are significant. According to
\citet{pp2004}:

\begin{equation}
12+\log\mbox{(O/H)} = 8.73 - 0.32 \times O3N2
\label{eq:O3N2}
\end{equation}

This relation holds for local H~II regions with $-1 < O3N2 < 1.9$
(or, according to the $O3N2$ calibration,
$8.12 < 12+\log(\mbox{O/H}) < 9.05$), and
arises from the fact that [OIII]/H$\beta$ decreases with increasing
metallicity, while, at least up to solar metallicity, [NII]/H$\alpha$
increases with increasing metal content. We derive oxygen abundances using
the $O3N2$ indicator for the five galaxies in our sample with
[OIII]/H$\beta$ and [NII]/H$\alpha$ measurements, as shown in
Table~2. The average value of
([OIII]/H$\beta$)/([NII]/H$\alpha$)  implies an oxygen abundance of
$12+\log\mbox{(O/H)}=8.38$. This value corresponds to $\sim 0.5
\mbox{(O/H)}_{\odot}$, when we compare with the solar value of
$12+\log\mbox{(O/H)}=8.66$ \citep{allende2002,asplund2004}.  The three
galaxies at $z\sim 1.4$ have systematically lower $O3N2$ values,
corresponding to higher oxygen abundances, than the two galaxies at $z\sim
1.0$.  However, since the two galaxies at $z\sim 1.0$ are less luminous,
the difference in average O/H between the two subsamples most likely
reflects the fundamental correlation between metallicity and luminosity,
rather than redshift evolution. A larger sample will be required to draw
further inferences.

Using the same H~II region calibration sample, 
\citet{pp2004} show, in terms of $N2 \equiv$~log([NII] 6584/H$\alpha$), that:

\begin{equation}
12+\log\mbox{(O/H)} = 8.90 + 0.57 \times N2
\label{eq:N2}
\end{equation}
The scatter around this relation is $\pm 0.18$~dex. It is valid
for H~II regions spanning from $-2.5 < N2 < -0.3$ (or, according
to the $N2$ calibration, from 
$7.50 < 12+\log(\mbox{O/H}) < 8.75$).
For the total sample
of 12 galaxies, including those with only [NII]/H$\alpha$, we calculate
O/H based on the $N2$ indicator alone. The results are listed in
Table~2. As shown with detailed photoionization models
\citep{kewley2002}, the $N2$ indicator is not sensitive to increasing
oxygen abundance above roughly solar metallicity, so is not an ideal
metallicity indicator. However, as long as we use the same empirical
quantities for samples at other redshifts, we can still use $N2$ to obtain
crude information from our total sample of 12 galaxies about the evolution
of fundamental scaling relationships between star-forming galaxy
luminosity and metallicity $(L-Z)$, and between mass and metallicity
$(M-Z)$. The average value of [NII]/H$\alpha$ for the sample is 0.25,
implying an oxygen abundance of $12+\log\mbox{(O/H)}=8.56$, which
corresponds to $\sim 0.8( \rm{O/H})_{\odot}$. In our total sample, the
three $z\sim 1.0$ galaxies have systematically lower [NII]/H$\alpha$
ratios than the $z\sim 1.4$ galaxies, but are also systematically fainter
-- consistent with the statistics in the smaller $O3N2$ sample.  
Comparing oxygen abundances derived from $N2$ versus those derived from
$O3N2$, for the five galaxies with both measurements, we find that the
$O3N2$-based oxygen abundances are systematically lower by
$0.10-0.15$~dex. The difference is slightly larger for the three galaxies
at $z\sim 1.4$, and may result from intrinsic differences in the H~II
region physical conditions in these galaxies. We will return to this point
in section~\ref{sec:bpt}.

\section{Luminosity-Metallicity and Mass-Metallicity Relationships}
\label{sec:lzmz}

Both star-forming and early-type galaxies in the local universe follow
strong correlations between rest-frame optical luminosity and the degree
of chemical enrichment \citep{garnett1987,brodie1991}.  Perhaps a more
fundamental correlation is the one observed between stellar mass and
metallicity. For star-forming galaxies, these correlations have recently
been measured over a factor of $10^3$ in stellar mass, 6 optical
magnitudes and a factor of 10 in gas-phase oxygen abundance, using a
sample of $>53,000$ non-AGN emission-line galaxies culled from the SDSS
\citep{tremonti2004}. While the oxygen abundance increases fairly rapidly
with increasing stellar mass at low mass, the relationship flattens at
stellar masses of $3\times10^{10} M_{\odot}$ and above.
\citet{tremonti2004} interpret this result as evidence for galactic winds
depleting the metal content of galaxies below a critical mass of
$3\times10^{10} M_{\odot}$, thereby lowering the associated effective
yield. The effect of outflows, commonly referred to as ``star-formation
feedback," is a crucial input to models of galaxy formation and may be
required as a fairly generic ingredient in order to explain a wide range
of observed phenomena, including the observed angular momentum in disk
galaxies, the metal content in the IGM, the global stellar content in the
local universe, and the amount of entropy in clusters. Measurement of the
redshift evolution in the observed relationships among metallicity,
luminosity, and stellar mass will provide powerful constraints on models
of galaxy formation, and, in particular, star formation feedback.

\subsection{Observed Evolution in the $L-Z$ and $M-Z$ Relationships}
\label{sec:lzmzobs}

We now construct the relationships among oxygen abundance, absolute
$B$-magnitude, and stellar mass for our sample of DEEP2 galaxies, and
compare with analogous relationships determined at lower redshift. Since
there is considerable scatter among the calibrations of commonly used
oxygen abundance indicators, a fair comparison of galaxy samples at
different redshifts must be based on the same empirical strong-line
indicator, using the same calibration. For a uniform comparison with the
$z\sim 1$ DEEP2 sample, we also apply the $N2$ and $O3N2$ indicators
\citep{pp2004} to the local SDSS comparison sample. These indicators
require measurement of the full set of [OIII], H$\beta$, H$\alpha$ and
[NII] for galaxies in each sample, all of which are available for the
$\sim 53,000$ SDSS emission-line galaxies with $\langle z \rangle = 0.1$,
presented in \citet{tremonti2004}.

Absolute blue magnitudes, $M_B$, are computed for SDSS galaxies based on
observed $g$-band magnitudes, with appropriate color- and $k$-corrections
to Johnson $B$ at $z=0$ \citep{tremonti2004}. For the three $z=1.0$ DEEP2
galaxies presented here, an analogous procedure can be applied, as the
observed $I$-band is a very good match to rest-frame $B$. However, for the
9 DEEP2 galaxies at $z\sim 1.4$, an extrapolation to observed $\sim 1
\mu$m is required. For both $z\sim 1.0$ and $z\sim 1.4$ objects, we use
the $M_B$ values from \citet{willmer2005} based on optical data alone, and
confirm their good agreement with estimates based on fits to the $BRIK_s$
SEDs that span through rest-frame $I$-band [$J$-band] for the $z\sim 1.4$
[$z\sim 1.0$] galaxies.

Stellar masses have been derived for SDSS galaxies, based both on spectral
features including the 4000\AA\ break and H$\delta_A$ Balmer absorption
index, and broadband photometry, as described in \citet{kauffmann2003}.
The library of star-formation histories used to fit the SEDs of SDSS
galaxies includes the possibility of past bursts as well as more smoothly
declining star-formation histories. We derive stellar masses for the 9
DEEP2 galaxies in our sample with $K$-band photometry, following the
procedure outlined in \citet{bundy2005}. This method uses the same general
Bayesian framework as the work of \citet{kauffmann2003}, though only
includes broadband photometry, and relatively simple star-formation
histories.  The observed $BRIK_s$ SED of each galaxy is compared with a
grid of synthetic SEDs from \citet{bc2003}, spanning a range of
single-component, exponentially-declining star-formation histories, ages,
and dust content. The $K$-band $M/L$ ratio, stellar mass, and probability
that the model represents the data are computed at each grid point. The
probabilities are marginalized across the grid and binned by stellar mass,
yielding a stellar mass probability distribution. We adopt the median and
standard deviation of this distribution as the estimated stellar mass and
uncertainty, respectively. To measure evolution in the relationship
between mass and metallicity, we must establish that there are no
systematic differences in the stellar masses derived using different
techniques. As a test, we model a subset of 500 SDSS galaxies using the
\citeauthor{bundy2005} technique, and find very good agreement -- the
average fractional stellar mass difference is $\langle
(M_{Bundy}-M_{SDSS})/M_{SDSS} \rangle = -0.03 \pm 0.32$. The scatter in
the average stellar mass difference amounts to a little less than a factor
of two, smaller than any of the evolutionary differences we report below.
More importantly, though, there is no significant systematic difference in
the two techniques used to model stellar masses. Finally, we note that
SDSS stellar masses are derived assuming a \citet{kroupa2001} stellar
initial mass function (IMF), while we assume the form of
\citet{chabrier2003}. These two IMF parametrizations yield roughly
consistent stellar masses for the same SED.

\begin{figure*}
\epsscale{1.0}
\plotone{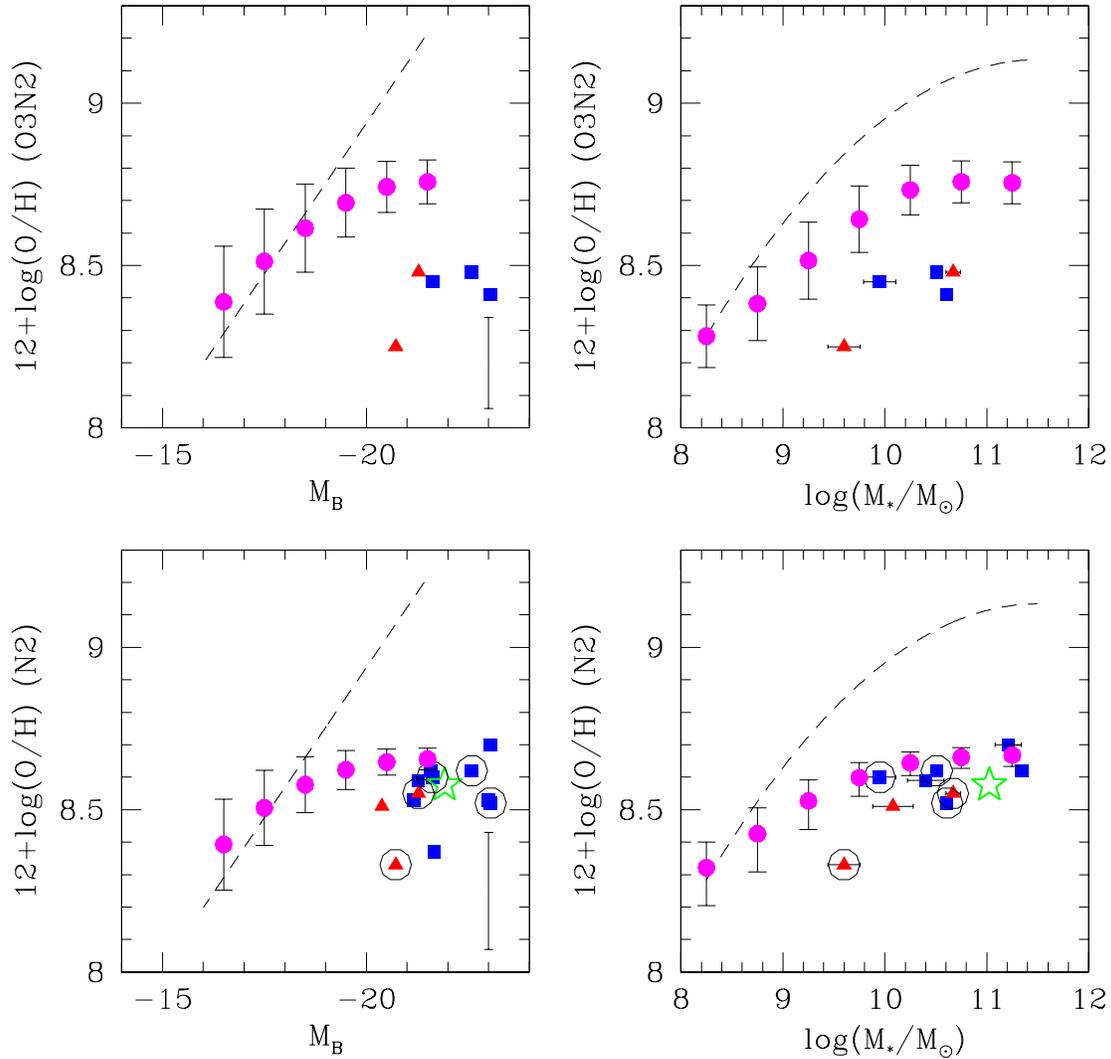}
\epsscale{1.0}
\caption{Luminosity-Metallicity (left) 
and Mass-Metallicity Relationships (right), based on $O3N2$ (top) and $N2$
(bottom). DEEP2 objects at $z\sim 1.0$ are indicated with red triangles,
while those at $z\sim 1. 4$ are indicated with blue squares. The five
DEEP2 objects with $O3N2$ measurements are circled in the panels showing
$N2$-based abundances. In each panel, average oxygen abundances based on
the $N2$ or $O3N2$ indicators are computed for SDSS objects, in bins of
$M_B$ and $M_*$. The SDSS average O/H values are indicated with the large
solid circles, with vertical error bars representing the 1$\sigma$ ranges
in the O/H values in each $M_B$ or $M_*$ bin. The saturation of the $N2$
indicator at high metallicities is readily apparent in the SDSS sample.
The $L-Z$ and $M-Z$ relationships based on the O/H values computed in
\citet{tremonti2004} are indicated as dashed lines. The systematic
differences between the dashed lines and solid circles -- i.e. the
differences resulting from using different indicators on the same galaxies
-- reveal the need for using the same abundance indicator when comparing
different galaxy samples. In the lower two panels, the average O/H, $M_B$,
and $M_*$ are indicated with an open star for the $z\sim 2$ sample of
\citet{shapley2004}. The vertical error bars in the lower right hand
corner of the $L-Z$ panels indicate the systematic uncertainty in the
$O3N2$ and $N2$ indicators.  Based on both $O3N2$ and $N2$, at fixed
oxygen abundance the DEEP2 galaxies are offset by several magnitudes
towards brighter luminosities and by a factor of $>10$ towards larger
stellar mass. Conversely, the DEEP2 galaxies are less metal-rich at fixed
luminosity and stellar mass.
}
\label{fig:lzmz}
\end{figure*}

Figure~\ref{fig:lzmz} shows the relationships among gas-phase oxygen
abundance, $B$-band luminosity, and stellar mass for galaxies at different
redshifts. In the top panels oxygen abundance is derived from the $O3N2$
indicator, whereas the $N2$ indicator is used in the lower panels.  As
described in section~\ref{sec:oh}, the $N2$ indicator saturates and is no
longer very sensitive to increasing O/H at [NII]/H$\alpha>0.3$. While only
25\% of the DEEP2 sample has [NII]/H$\alpha > 0.3$, more than 75\% of the
SDSS sample of \citet{tremonti2004} has line ratios in this regime. Given
the luminosity and stellar mass ranges we are considering here, the
$N2$-based plots are less sensitive to differences in metallicity in the
SDSS sample at fixed luminosity or stellar mass. However, including
$N2$-based abundances allows us to use our full sample of 12 DEEP2
objects, which span a larger range in luminosity and stellar mass, as well
as the sample of $z\sim 2$ galaxies presented in \citet{shapley2004}. The
SDSS data have been binned in intervals of $\Delta M_B=1.0$ and $\Delta
M_*=0.5$, with the average value $12+\log\mbox{O/H}$ shown for each bin in
luminosity and stellar mass. To demonstrate the importance of using the
same empirical abundance indicator for each sample, we also plot the
best-fit $L-Z$ and $M-Z$ relations from \citet{tremonti2004}. The O/H
values used to derive these relationships come from models of the full set
of strong line spanning in wavelength from [OII] to [SII], and are
systematically higher than the $O3N2$ and $N2$-based O/H abundances. The
difference increases at increasing O/H.

Our current sample of $z\sim 1.0-1.5$ galaxies is too small to determine
the detailed form of trends among metallicity, luminosity and stellar mass
in the manner of \citet{tremonti2004}. However, when the total sample of
12 [NII]/H$\alpha$ measurements is divided into faint and bright
subsamples, we find a higher average [NII]/H$\alpha$ ratio for the bright
subsample, a roughly one sigma difference. Dividing the sample into low
and high stellar mass bins also indicates a higher [NII]/H$\alpha$ ratio
as a function of increasing stellar mass. Finally, we note that the
$z=1.04$ galaxy, 32020728, is the faintest and least massive object among
the sample of galaxies with $O3N2$ measurements, and has the lowest
inferred O/H abundance as well. In the future, it will be important to
probe a large sample of objects over as large a magnitude and stellar mass
range as possible, in order to confirm the nature of the trends that we
have begun to trace with our small pilot sample. Despite the small size of
our distant galaxy sample, we still discern significant evolution in at
least the zeropoints of the relationships among metallicity, luminosity
and stellar mass, as presented below. This evolution in $L-Z$ and $M-Z$
space can be used to understand how the $z\sim 1$ star-forming galaxies
relate to star-forming galaxies in the nearby universe.

First we consider evolution in the relationship between O/H and $M_B$.
Regardless of the indicator used, DEEP2 galaxies at $z\sim 1$ are several
magnitudes brighter than $\langle z \rangle=0.1$ SDSS galaxies at fixed
O/H.  For a more meaningful comparison with SDSS galaxies, we use the
subsample of DEEP2 galaxies with $O3N2$-based abundances. These galaxies
have $\langle M_B \rangle = -21.9$ and $\langle 12+\log(\mbox{O/H})
\rangle = 8.4$. SDSS galaxies with comparable $12+\log(\mbox{O/H})$ have
$\langle M_B \rangle = -16.7$, 5.2 magnitudes fainter!  SDSS galaxies with
comparable $M_B$ to the average of the DEEP2 sample have $\langle
12+\log(\mbox{O/H}) \rangle = 8.75$, 0.35 dex higher.  Now we turn to the
question of stellar mass. The DEEP2 sample with $O3N2$ measurements have
$\langle M_* \rangle = 3\times 10^{10} M_{\odot}$. SDSS galaxies with
comparable $12+\log(\mbox{O/H})$ have $\langle M_* \rangle = 8\times
10^{8} M_{\odot}$. SDSS galaxies with comparable stellar mass to the
average of the DEEP2 sample have $\langle 12+\log(\mbox{O/H}) \rangle =
8.7$. While SDSS contains a significant fraction of galaxies with higher
masses than the average among the DEEP2 sample, the inferred O/H does not
significantly increase at stellar masses larger than $10^{10} M_{\odot}$.
The flattening in the mass-metallicity relationship at high stellar mass
reflects the physical properties of the galaxies \citep{tremonti2004},
though the absolute value of O/H at which the trend flattens differs
depending on which abundance indicator is used. According to the abundance
scale adopted in \citet{tremonti2004}, the local M-Z relationship flattens
at $12+\log(\mbox{O/H}) \sim 9.1$, whereas the $O3N2$-based scale
indicates flattening at $12+\log(\mbox{O/H}) \sim 8.8$. This difference
reflects the magnitude of systematic uncertainties among various
strong-line calibrations in the high-metallicity regime.

Several groups have considered the evolution of the $L-Z$ relationship at
$0 < z < 1$, primarily based on $R_{23}$-based oxygen abundances. The
largest sample of intermediate redshift galaxies with oxygen abundance
measurements is contained in \citet{kobulnicky2004}, with 204
emission-line galaxies at $0.30 < z < 0.94$ drawn from the Team Keck
Redshift Survey (TKRS). Of these, 38 galaxies with $\langle z \rangle
=0.4$ have both [OIII]/H$\beta$ and [NII]/H$\alpha$ measurements, allowing
a direct comparison with our DEEP2 sample. The TKRS subsample with
[NII]/H$\alpha$ measurements is roughly equidistant in lookback time
between the SDSS and DEEP2 samples, and contains $M_B$ values at fixed O/H
that are intermediate between those of our DEEP2 sample and SDSS galaxies.
Specifically, galaxies in the $\langle z \rangle =0.4$ TKRS subsample with
$12+\log(\mbox{O/H}) \sim 8.4$ have $\langle M_B \rangle = -20.2$. Using
$R_{23}$ for the total TKRS sample at $0.3 \leq z < 0.94$,
\citet{kobulnicky2004} find that the intermediate-redshift galaxies are
$1-3$ magnitudes more luminous at a given metallicity than their local
counterparts. The evolution in the $L-Z$ relationship at intermediate
redshift is not resolved at the very luminous end, however, as both
\citet{lilly2003} and \citet{kobulnicky2003} find little difference in the
average metallicities of the most luminous ($M_B<-20$) galaxies in their
samples -- even at $0.6 \leq z < 1.0$ -- relative to local counterparts.
Larger samples at both intermediate redshift and $z\geq 1$ will help
resolve this ambiguity.  For now, though, we focus on the consistent
picture of evolution provided by our DEEP2 sample, the TKRS galaxies with
[NII]/H$\alpha$ measurements, and the SDSS emission-line sample.

\subsection{The Evolution of DEEP2 Galaxies}
\label{sec:lzmzevol}

In order to explain the evolution in the $L-Z$ and $M-Z$ relationships
observed for DEEP2 galaxies at $z\geq 1$ and SDSS galaxies at $z\sim 0.1$,
we make two key assumptions. First, we assume that the descendants of the
DEEP2 galaxies with abundance measurements will be forming stars at $z\sim
0.1$. Accordingly, they should be included in the emission-line sample of
\citet{tremonti2004}, and follow the $L-Z$ and $M-Z$ relationships
observed among SDSS galaxies. This is a reasonable assumption, based on
the clustering strength of blue DEEP2 galaxies with similar spectral and
photometric properties. \citet{coil2004} find that DEEP2 galaxies with
strong emission lines and also those with blue rest-frame optical colors
have correlation lengths of $r_0 \sim 3h^{-1}$ comoving Mpc. The local
descendants of the dark matter halos hosting such galaxies will cluster
more like blue, emission-line galaxies in the \citeauthor{tremonti2004}
SDSS sample, rather than red, absorption-dominated galaxies
\citep{zehavi2005,budavari2003,adelberger2005a}. Also, based on the
relative evolution of the rest-frame $B$-band luminosity functions of red
and blue galaxies between $z\sim 1$ and $z\sim 0$, it is reasonable to
assume that a significant fraction of the blue DEEP2 galaxies will remain
blue \citep{faber2005}. Second, we assume that DEEP2 galaxies fade on
average by $\Delta(M_B)=1.3$~mags between $z\sim 1$ and $z\sim 0$. This
degree of fading reflects the observed evolution of $M_B*$ in both blue
and red galaxy luminosity functions traced to $z\geq 1$ by the DEEP2 and
COMBO-17 surveys \citep{willmer2005,faber2005}.

Following these two assumptions, we shift the DEEP2 sample in luminosity
by an amount that reflects the average evolution $M_B*$ between $z\sim 1$
and $z\sim0$, $1.3$~mag. This shift implies $\langle M_B \rangle = -20.55$
at $z\sim 0$ for the DEEP2 galaxy descendants. In order for these local
counterparts to lie on the local $L-Z$ relationship, the sample average
oxygen abundance must increase by 0.35~dex to $12+\log(\mbox{O/H})=8.75$.
SDSS galaxies with $M_B\sim -20.55$ have average stellar masses of
$\langle M_*\rangle \sim 4\times 10^{10} M_{\odot}$, twice the average
stellar mass of our DEEP2 sample. Therefore, these possible DEEP2
descendants have stellar $M/L_B$ ratios a factor of $6-7$ higher on
average than those at $z\sim 1$.

It is instructive to compare the concurrent increases in O/H and stellar
mass, described above, with the predictions of chemical evolution models.
According to a simple ``closed box'' model of chemical evolution, the
increase in gas-phase metallicity is governed by the fraction of gas that
has been converted into stars, and the yield of metals returned to the ISM
by each generation of stars (under the assumption of instantaneous
recycling). The model is described by the equation:  

\begin{equation} 
Z = y \ln(1/\mu_{gas}) 
\end{equation}

where $Z$ is the mass fraction of metals in the ISM, $y$ is the metal
yield by mass, and $\mu_{gas}$ is the fraction of the total baryonic mass
remaining in gas that has not yet been converted into stars. Predicting
the increase in stellar mass for a given increase in O/H requires an
estimate of the yield of oxygen. Based on the solar neighborhood value of
$Z_{\odot}=0.02$, and with oxygen comprising $\sim 45$\% of the mass in
metals, past estimates of the oxygen yield by mass were $y_O=0.009$
\citep{lee2003}. New measurements of the solar abundance may indicate that
$y_O$ is actually $\sim 0.2$ dex lower. In an independent manner, both
\citet{garnett2002} and \citet{tremonti2004} compute the ``effective
yield'', $y_{eff}$, for star-forming galaxies as a function of mass. The
quantity $y_{eff}$ is defined as the oxygen abundance divided by
$\ln(1/\mu_{gas})$.  Both authors find that $y_{eff}$ asymptotes at large
galaxy masses to a value of $y_O=0.01$, which is considered the true
yield. However, the value of $y_O$ derived by both \citet{garnett2002} and
\citet{tremonti2004} is derived from O/H values based on the SDSS or
$R_{23}$ abundance scales, both of which yield systematically higher
values than $O3N2$ calibration.  With the $O3N2$ calibration, the value of
the true yield could easily be lower by $\sim 0.3$ dex.

Given the uncertainty in $y_O$, which could range between $y_O=0.006$ and
$0.01$, we would predict an increase in stellar mass of a factor of
$1.7-1.9$ to accompany the increase in metallicity of DEEP2 galaxies
between $z\sim 1$ and $z\sim 0$, if the systems are treated as closed
boxes. Therefore, the observed increase in stellar mass of roughly a
factor of two is consistent with expectations of closed box chemical
evolution models, for an increase in average oxygen abundance from
$12+\log(\mbox{O/H})=8.40$ to 8.75. Approximate closed boxes may not be
such unrealistic descriptions of these galaxies, at least in terms of the
importance of outflows.  The likely SDSS descendants of the DEEP2 galaxies
have effective yields, derived by \citet{tremonti2004}, that are within
only $0.1$~dex of the true yield, which is interpreted as an indication
that the effect of supernova feedback cannot be significant on the
progress of enrichment in these galaxies.

\subsection{A Comparison with Star-forming Galaxies at $z\sim 2$}
\label{sec:lzmzbx}

[NII]/H$\alpha$ ratios have also been measured for star-forming galaxies
at $z\sim 2$ \citep{shapley2004}. Measurements for individual galaxies are
available for $z\sim 2$ objects with $K_s\leq 20.0$, the most luminous
$\sim 10$\% (in the rest-frame optical) of UV-selected galaxies in this
redshift range \citep{steidel2004}. [NII]/H$\alpha$ can be measured from
composite near-IR spectra of fainter galaxies in the sample as well. As
shown by the open star in the lower panels of figure~\ref{fig:lzmz},
representing the average $12+\log(\mbox{O/H})$, $M_B$, and $M_*$ for the
sample of 7 $K_s\leq 20.0$ $z\sim 2$ galaxies in \citet{shapley2004}, the
average $M_B$ and O/H values for the sample are very similar to those of
the DEEP2 sample. Furthermore, \citet{shapley2004} show preliminary
evidence for an $L-Z$ relationship among $z\sim 2$ star-forming galaxies,
by demonstrating that the [NII]/H$\alpha$ ratio is smaller in a composite
spectrum of a sample of $z\sim 2$ galaxies with fainter rest-frame optical
luminosities \citep{erb2003}.  Despite similar locations in the $L-Z$
plane, the $z\sim 2$ galaxies in \citet{shapley2004} will follow
significantly different evolutionary paths relative to those of our DEEP2
sample. Specifically, the $z\sim 2$ sample will not be forming stars by
$z\sim 0$, therefore their local descendants will not be found among the
sample of \citet{tremonti2004}. In fact, it is likely that they will no
longer be forming stars even by $z\sim 1$, and will be found among the
red, absorption-line galaxies in the DEEP2 survey.

This scenario is supported by multiple pieces of evidence. First, the
average stellar mass and $M/L_B$ ratio among the higher-redshift
\citet{shapley2004} sample is already $\sim 3$ times higher than that of
the DEEP2 sample average. This difference is shown graphically by the the
location of the star symbol in figure~\ref{fig:lzmz}, representing the
average of the $z\sim 2$ sample. While this point lies squarely in the
middle of the distribution of DEEP2 $M_B$ values (left-hand panel), it is
found towards the high-mass extreme of the DEEP2 stellar mass values
(right-hand panel).  $M/L_B$ ratios will generally tend to increase with
time, barring the presence of a major later episode of star-formation.
Only one galaxy in our DEEP2 $N2$ sample has a stellar $M/L_B$ greater
than those in the $z\sim 2$ comparison sample, and may serve as a $z\sim
1$ counterpart. The SED of this one object, 32025514, indicates the
highest stellar mass in the total sample of 12 DEEP2 galaxies, and a
significantly more mature stellar population as traced empirically by the
largest $R-K_s$ color (rest-frame UV-to-optical break). In contrast to
most of the galaxies in our DEEP2 sample, 32025514 may follow an
evolutionary path between $z\sim 1$ and $z\sim 0$ resulting in a local
descendent with no ongoing star formation. In the future, it will be
valuable to compare the DEEP2 sample with a more representative sample of
$z\sim 2$ objects, spanning a larger dynamic range in stellar mass and
$M/L_B$ (Erb et al., in preparation).

Another clue about the evolution of galaxies in the \citet{shapley2004}
sample is based the large clustering strength of UV-selected galaxies at
$z\sim 2$. The correlation length among the total $z\sim 2$ sample is
$4.2\pm 0.5 h^{-1}$ comoving Mpc, and among galaxies with $K_s<20.5$, it
increases to $10\pm 3 h^{-1}$ comoving Mpc
\citep{adelberger2005a,adelberger2005b}. A comparison with simulations
indicates that red, absorption-line dominated galaxies in the DEEP2 survey
cluster as strongly as the $z\sim 1$ descendants of the dark matter haloes
hosting typical UV-selected galaxies at $z\sim 2$; elliptical galaxies in
the SDSS cluster like the $z\sim 0$ descendants of the same dark matter
haloes \citep{coil2004,budavari2003,adelberger2005a}. The correlations
observed in $z\sim 2$ star-forming galaxies among O/H, $M_B$, and $M_*$
can therefore be used to understand the formation of elliptical galaxies.
As the \citet{shapley2004} sample all have $K_s \leq 20.0$, their
descendants at $z\sim 1$ and $z\sim 0$ should be more massive, more
strongly clustered, and even less likely to be forming stars than the
descendants of typical UV-selected $z\sim 2$ galaxies.

\section{H~II Region Physical Conditions}
\label{sec:bpt}

\subsection{Emission-Line Diagnostics}
\label{sec:bptem}

\begin{figure*}
\epsscale{1.0}
\plotone{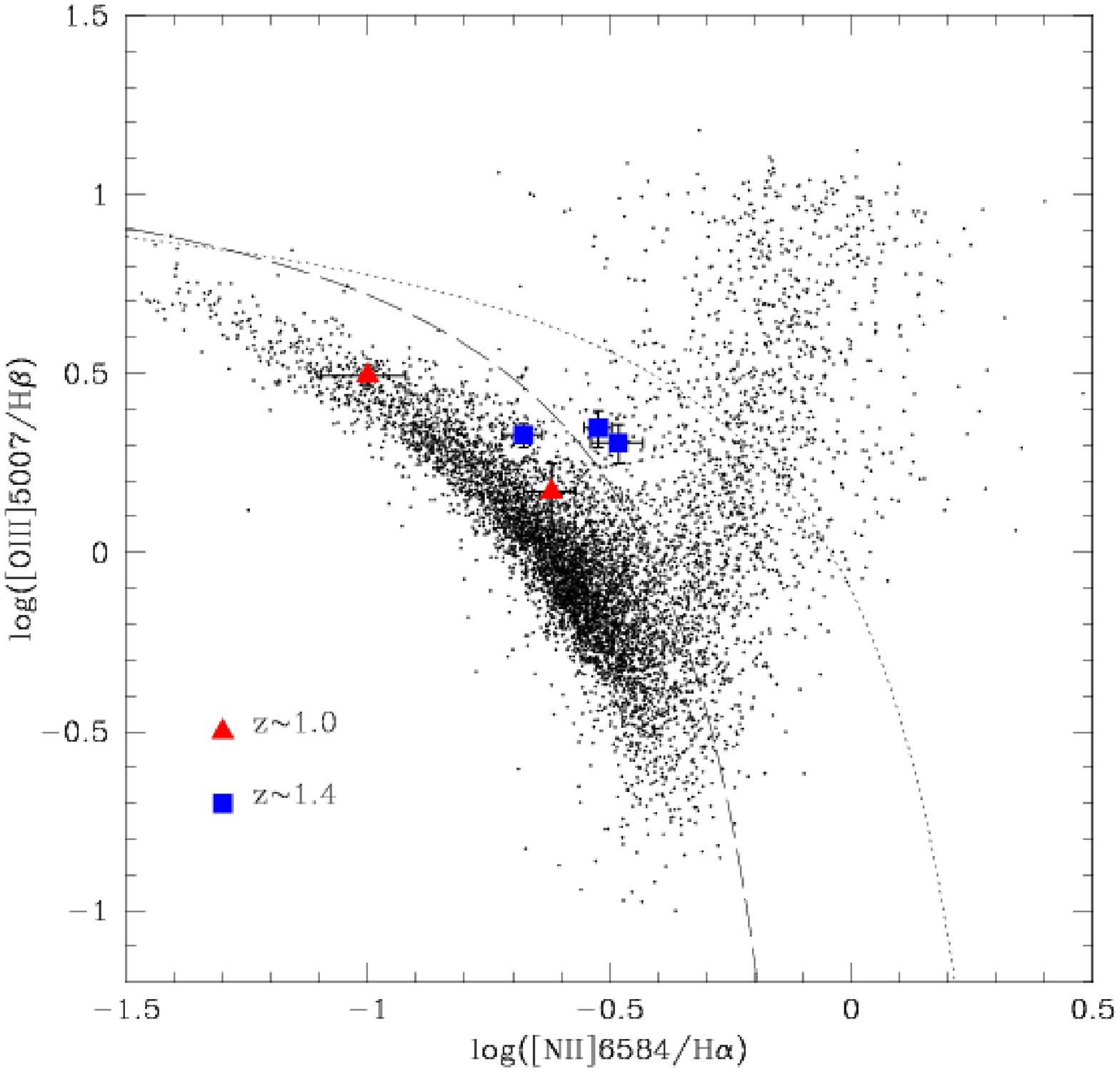}
\epsscale{1.0}
\caption{H~II region Diagnostic Diagram. 
$z\sim 1.0$ DEEP2 galaxies are indicated as large red triangles, and
$z\sim 1.4$ galaxies as large blue squares. Small black points are
galaxies from the SDSS \citep{tremonti2004,brinchmann2004}. This plot is
used to distinguish star-forming galaxies from AGN.  The dashed line is an
empirical demarcation from \citet{kauffmann2003b} based on the SDSS
sample, whereas the dotted line is the theoretical limit for star-forming
galaxies from \citet{kewley2001}, based on detailed photo-ionization plus
stellar population synthesis models. Star-forming galaxies and H~II
regions form a well-defined excitation sequence of photo-ionization by hot
stars, below and to the left of the dashed and dotted curves. LINERs and
AGN lie above and to the right, with emission line ratios reflecting
photo-ionization by a non-thermal power-law or radiative shocks.  The
DEEP2 objects at $z\sim 1.4$ are significantly offset from the excitation
sequence followed by SDSS galaxies and nearby H~II regions. These distant
galaxies are located in a region of parameter space intermediate between
typical local star-forming galaxies and AGN, though below the theoretical
starburst limit of \citet{kewley2001}.
}
\label{fig:o3n2}
\end{figure*}

In Figure~\ref{fig:o3n2}, [OIII]/H$\beta$ and [NII]/H$\alpha$ ratios are
plotted for the five galaxies with the full set of emission lines. This
space serves as a fairly standard diagnostic for discriminating between
star-forming galaxies and AGN \citep{bpt1981,veilleux1987}.  Also
indicated in Figure~\ref{fig:o3n2} are emission-line objects from the SDSS
survey, which trace out well-defined sequences. Star-forming galaxies
occupy the fan in the lower left-hand portion of the diagram, in a region
roughly defined by decreasing excitation as a function of increasing
metallicity. This physical sequence results in the empirical
anti-correlation between [OIII]/H$\beta$ and [NII]/H$\alpha$ up to the
point where [NII]/H$\alpha$ saturates (at [NII]/H$\alpha\sim 0.3$). At the
highest metallicities, [OIII]/H$\beta$ continues to decrease at relatively
fixed [NII]/H$\alpha$. AGN mostly occupy the upper right-hand region of
the diagram, typically described by both higher [NII]/H$\alpha$ and
[OIII]/H$\beta$ ratios than those in star-forming galaxies. To place the
empirical H~II region sequence on the theoretical footing of ionizing
spectrum, metallicity, and ionization parameter, most recently
\citet{dopita2000} used PEGASE \citep{fioc1997} and STARBURST99
\citep{leitherer1999} instantaneous zero-age stellar population models for
the input ionizing spectra, metallicities ranging from $Z=0.05-3.0
Z_{\odot}$ with the \citet{anders1989} values of solar metallicity,
ionization parameters ranging from $q=5\times 10^{6}$ to $3\times10^{8}
\mbox{ photons}\mbox{ cm}\mbox{ s}^{-1}$, and an electron density of
$n_e=10 \mbox{ cm}^{-3}$. With these models, \citet{dopita2000}
successfully recovered the location and upper envelope of the
[OIII]/H$\beta$ vs. [NII]/H$\alpha$ emission-line sequence of local H~II
regions, which is observed among both the H~II regions compiled by
\citet{pp2004} to generate the $O3N2$ abundance calibration, and the
emission-line galaxies in \citet{tremonti2004}. Two curves are plotted in
Figure~\ref{fig:o3n2}, of the same shape as the upper-envelope found by
\citet{dopita2000}, though with different normalizations. These curves
serve to discriminate between star-forming galaxies and AGN. The more
conservative, dashed curve is from \citet{kauffmann2003b}, and is used to
define the sample of star-forming galaxies presented in
\citet{tremonti2004}. This criterion should yield galaxies in which the
H$\alpha$ flux has $< 1$\% contribution from AGN \citep{brinchmann2004}.
The upper dotted curve is from \citet{kewley2001} and represents a
theoretical upper limit on the location of star-forming galaxies in the
diagnostic diagram.  It is very unlikely to find star-forming galaxies
above the line of \citet{kewley2001}.  We shall return to this limit in
the discussion below.

While the two $z\sim 1.0$ galaxies in our DEEP2 sample are roughly
consistent with the sequence of SDSS star-forming galaxies and local H~II
regions in terms of diagnostic line ratios, the three $z\sim 1.4$ galaxies
are significantly offset from this sequence. The difference between $z\sim
1.0$ and $z\sim 1.4$ galaxies more likely reflects the nearly disjoint
ranges of rest-frame luminosity and star-formation rate probed by the
small samples, rather than an evolutionary trend within DEEP2. Two of the
$z\sim 1.4$ galaxies, 42008627 and 42008219, are offset towards higher
[OIII]/H$\beta$ by $0.8$~dex, relative to star-forming SDSS galaxies with
similar [NII]/H$\alpha$ ratios. The third galaxy, 42055947, is offset by
0.3~dex, which is still significant. Conversely, 42008627 and 42008219 are
displaced by 0.4~dex towards higher [NII]/H$\alpha$ relative to SDSS
galaxies with similar [OIII]/H$\beta$ ratios, while 42055947 is displaced
by 0.2~dex. The line ratios of 42008627 and 42008219 place these objects
in a region of the diagnostic diagram only sparsely populated by SDSS
emission-line galaxies, intermediate between local star-forming galaxies
and AGN, above the \citet{kauffmann2003b} curve but below that of
\citet{kewley2001}.  In a preliminary sample of four UV-selected $z\sim 2$
star-forming galaxies with the full set of emission-line measurements,
offsets of equal or greater value relative to the emission-line sequence
of local star-forming galaxies have been measured, and over a larger range
in [NII]/H$\alpha$ (Erb et al., in preparation). The subset of the
intermediate redshift TKRS sample \citep{kobulnicky2004} with measured
[NII]/H$\alpha$ ratios follows an emission-line sequence consistent with
the SDSS sequence over most of the range in [NII]/H$\alpha$. At
[NII]/H$\alpha>0.25$, however, there are a few outlying TKRS galaxies with
similar line ratios to those of the $z\sim 1.4$ DEEP2 galaxies.

One effect that could potentially bias the measured [OIII]/H$\beta$ ratios
towards higher values is the underestimate of H$\beta$ due to underlying
stellar Balmer absorption.  Stellar absorption will have less of an effect
on H$\alpha$ because of its larger emission line strength. The effect of
stellar absorption was not included in the calculation of [OIII]/H$\beta$
and [NII]/H$\alpha$ ratios in part because little or no continuum was
detected. However, it is possible to place an upper limit on the
importance of this effect for the $z\sim 1.4$ galaxies with offset line
ratios by using the $3\sigma$ upper limit on the continuum,
$F_{\lambda,3\sigma}$, and a reasonable estimate of the stellar absorption
equivalent width, $EW(\mbox{H}\beta)=5$~\AA\ \citep{charlot2002}.  The
upper limit on the amount by which the H$\beta$ flux must be corrected is
then $\Delta(F)=F_{\lambda,3\sigma} \times EW(\mbox{H}\beta)$. With such
corrections, the points for the $z\sim 1.4$ DEEP2 galaxies would shift by
no more than 0.1~dex downwards, and by an insignificant amount in
[NII]/H$\alpha$. Therefore, unaccounted-for stellar Balmer absorption is
not the explanation for the offsets in emission line ratios. With the
assumption that the offsets correspond to a difference in the intrinsic
emission line properties of the DEEP2 galaxies, we consider possible
causes and implications of this difference in the following discussion.

\subsection{Physical Differences in the H~II Regions of DEEP2 Galaxies}
\label{sec:bptphys}

First, it is worth placing the inferred star-formation rates of our DEEP2
sample in context with those of other distant star-forming galaxies, as
well as typical SDSS emission-line galaxies and intermediate-redshift TKRS
galaxies. The three $z\sim 1.4$ galaxies offset from the SDSS
emission-line diagnostic locus have inferred H$\alpha$ star-formation
rates ranging between $11-16 M_{\odot}\mbox{ yr}^{-1}$, uncorrected for
dust-extinction, while the two $z\sim 1.0$ galaxies have star-formation
rates of $\sim 5 M_{\odot}\mbox{ yr}^{-1}$. \citet{pettini2001} found an
average star-formation rate of $40 M_{\odot}\mbox{ yr}^{-1}$ for a sample
of 14 $z\sim 3$ Lyman Break Galaxies (LBGs). These star-formation rates
were based on H$\beta$ emission fluxes, since H$\alpha$ is shifted into
the thermal infrared at $z\sim 3$. For a sample of 16 galaxies at $z\sim
2.3$, selected with criteria designed to find $z\sim 2$ analogs to LBGs,
\citet{erb2003} found an average H$\alpha$ star-formation rate of $16
M_{\odot}\mbox{ yr}^{-1}$. The \citet{erb2003} sample therefore features a
comparable range of H$\alpha$ star-formation rates to that of the $z\sim
1.4$ DEEP2 galaxies.

The average H$\alpha$ star-formation rate in the SDSS sample, based on the
H$\alpha$ fluxes and redshifts in the sample of \citet{tremonti2004}, is
$0.5 M_{\odot}\mbox{ yr}^{-1}$ (the median is $0.3 M_{\odot}\mbox{
yr}^{-1}$). We correct this value by the 24\% median fraction of galaxy
light in the SDSS fiber apertures reported by \citet{tremonti2004}, which
translates into an aperture-corrected mean star-formation rate of $2.0
M_{\odot}\mbox{ yr}^{-1}$ (and a median of $1.2 M_{\odot}\mbox{
yr}^{-1}$).  These values are almost an order of magnitude smaller than
the star-formation rates probed by our DEEP2 sample at $z\sim 1.4$.  
Furthermore, only 1.5\% of the \citet{tremonti2004} sample has
aperture-corrected H$\alpha$ star-formation rates above $11
M_{\odot}\mbox{ yr}^{-1}$, in the range probed by our DEEP2 sample. We
find that SDSS galaxies with such high star-formation rates are indeed
offset towards higher [OIII]/H$\beta$ and [NII]/H$\alpha$ relative to the
\citet{tremonti2004} sample as a whole. The offset is not as significant
as seen in our DEEP2 sample at $z\sim 1.4$, though, because, by
definition, the \citet{tremonti2004} sample only contains galaxies below
and to the left of the \citet{kauffmann2003b} curve to insure minimal
contamination by AGN. As a caveat, we mention that dust extinction
typically amounts to a factor of $\sim 2$ correction for SDSS H$\alpha$
star-formation rates; but, as no robust extinction correction was
available for the DEEP2 galaxies, we refrain from correcting the SDSS
star-formation rates at this point.  The $\langle z \rangle = 0.4$ TKRS
sample with [NII]/H$\alpha$ measurements \citep{kobulnicky2004} has an
average star-formation rate of $0.9 M_{\odot}\mbox{yr}^{-1}$, based on
H$\beta$ fluxes uncorrected for dust extinction. Furthermore, the
star-formation rates in this sample are rather modest, with all of them $<
4 M_{\odot}\mbox{yr}^{-1}$.

In summary, the DEEP2 galaxies are hosting significantly more active
levels of star-formation than the bulk of galaxies in the SDSS sample of
\citet{tremonti2004} and the $\langle z \rangle = 0.4$ TKRS subsample of
\citet{kobulnicky2004}.  It is therefore not surprising that their
emission-line diagnostics more closely resemble those of the nearby
luminous starbursts discussed below \citep{kewley2001b}, and of actively
star-forming galaxies at $z\sim 2$ \citep{erb2003}, rather than those of
H~II regions in more quiescent galaxies.  We now consider possible causes
of the offset in line ratios observed in distant star-forming galaxies.
These include differences in the ionizing spectrum, ionization parameter,
and H~II region electron density, the effects of shock ionization, and
contributions to the emission from an AGN.

An offset upwards and to the right of the local H~II region emission-line
sequence was previously found by \citet{kewley2001,kewley2001b} for a
sample of 157 local starburst galaxies selected by their warm infrared
colors. The starbursts in the \citet{kewley2001b} sample not accounted for
by the upper envelope of the H~II region emission-line sequence of
\citet{dopita2000} are found in a similar location to the DEEP2 $z\sim
1.4$ galaxies, in [OIII]/H$\beta$ vs. [NII]/H$\alpha$ parameter space.
\citet{kewley2001} find that no combination of ionization parameter and
metallicity can account for all the objects in their sample, if a zero-age
instantaneous burst of star formation is assumed for the input ionizing
spectrum. They infer that a harder extreme ultraviolet (EUV) ionizing
spectrum in the $1-4$~Rydberg range (i.e., between the H~I and He~II
ionization edges) is required to explain the offset in line diagnostics.
Such an EUV radiation field is obtained in \citet{kewley2001} by assuming
PEGASE continuous star formation models with an age of 6~Myr (at which
point an equilibrium is reached in the ionizing spectrum). These models
include fairly hard Wolf-Rayet star spectra in the $1-4$~ Rydberg range,
and the upper envelope of the associated ionization-parameter/metallicity
grid is what forms the \citet{kewley2001} upper limit curve for starbursts
in the [OIII]/H$\beta$ vs. [NII]/H$\alpha$ space (Figure~\ref{fig:o3n2}).
While the continuous PEGASE model can account for line ratios of the warm
starbursts in \citet{kewley2001} as well as those of the $z\sim 1.4$ DEEP2
galaxies, the Wolf-Rayet stellar atmospheres used by PEGASE to generate
such a hard EUV radiation field are not very realistic \citep{kewley2001}.
Therefore, this specific model may not represent the ultimate or only
solution to the problem of the offset line ratios. For example, the
spectral slope of the EUV spectrum in the $1-4$~Ryd range is also affected
by the slope of the stellar IMF. An IMF slope flatter than the standard
Salpeter value of $\alpha=-2.35$ will result in a higher fraction of the
most massive stars, with an accordingly harder ionizing spectrum
\citep{leitherer1999}.  At present, we have no independent observational
constraints on the slope of the high-mass stellar end of the IMF in DEEP2
galaxies.

There are other possible causes of the offset in the DEEP2 line ratios.  
As discussed above, our sample of DEEP2 galaxies at $z\sim 1.4$ hosts
elevated levels of star formation, compared with average galaxies in the
SDSS emission-line sample of \citet{tremonti2004}. Ionization parameter
and star-formation rate are not directly proportional -- since ionization
parameter indicates the ratio of densities of ionizing photons and
electrons, it is related to the star-formation rate through a geometrical
factor and through the electron density. However, elevated rates of
star-formation do provide a larger reservoir of ionizing photons. In the
sample of \citet{pettini2001}, where the inferred star-formation rates are
even more extreme than in the DEEP2 sample, large measured [OIII]/[OII]
line ratios indicate significantly higher ionization parameters than those
observed in more quiescent samples \citep{lilly2003}.  A robust extinction
correction based on simultaneously-measured H$\alpha$ and H$\beta$ fluxes
would be required to estimate an extinction-corrected [OIII]/[OII] value
and therefore an ionization parameter. Our current data do not satisfy all
of these requirements, but we will attempt to address the issue with
future observations. Furthermore, it will be possible to estimate the
ionization parameters for the small sample of SDSS galaxies found in the
same region of [OIII]/H$\beta$ vs. [NII]/H$\alpha$ space as the DEEP2
galaxies.

Another apparent physical difference between the DEEP2 star-forming
galaxies and the SDSS sample is found in terms of the typical H~II region
electron density. In the SDSS sample of \citet{tremonti2004}, the median
[SII]$\lambda6717$/[SII]$\lambda6731$ line ratio implies an electron
density of $\sim 50 \mbox{ cm}^{-3}$.\footnote{For SDSS spectra, which are
characterized by $\sim 2$\AA\ resolution, the more widely-spaced [SII]
line ratio is preferable to the [OII] line ratio for measuring electron
densities.} Since our NIRSPEC spectra are not deep enough to detect [SII]
for individual objects, we estimate electron densities for the five DEEP2
galaxies with $O3N2$ values based on the density-sensitive
[OII]$\lambda3729$/[OII]$\lambda3726$ line ratio found in the DEIMOS
discovery spectra.  While the DEIMOS [OII] spectra have not been carefully
flux-calibrated, the [OII] line ratios should be fairly insensitive to
flux-calibration. We find that the median [OII] line ratio implies an
electron density of $\sim 400\mbox{ cm}^{-3}$, which is significantly
higher than the typical value observed in SDSS emission-line galaxies. On
the other hand, the DEEP2 electron density is quite similar to the
electron density found in nearby starbursts. \citet{kewley2001b} report an
average electron density of $350\mbox{ cm}^{-3}$, based on [SII] line
ratios for their sample of warm infrared starbursts. This value is
consistent with other recent results for electron densities in starburst
nuclear H~II regions
\citep{ravindranath2001,contini1998,storchi1995,heckman1990}, though
higher than observed in typical disk H~II regions \citep{zaritsky1994}.
Higher electron densities in starburst nuclei result from the higher
ambient interstellar pressures present in such environments
\citep{dopita2005}.  For fixed ionization parameter, metallicity, and
input ionizing spectrum, the photo-ionization models presented in
\citet{kewley2001} display a dependence on electron density. The
dependence is in the sense that model grids with higher electron density
have an upper envelope in the space of [OIII]/H$\beta$ vs. [NII]/H$\alpha$
that is offset upwards and to the right, relative to model grids with
lower electron density. Therefore, the enhanced electron densities in
environments hosting active star-formation may account for some of the
differences in observed line ratios.

The next possibility to consider is the effect of shocks on the emergent
emission-line spectrum. Given the high star-formation rates observed in
the DEEP2 galaxies, such shock excitation might be produced by the
combined effects of supernovae explosions and stellar winds that
collectively result in the large-scale outflows observed in both local and
distant starbursts \citep{heckman1990,pettini2001}. Models for fast,
radiative shocks by \citet{dopita1995} predict a range of [OIII]/H$\beta$
and [NII]/H$\alpha$ line ratios that encompasses those observed for the
$z\sim 1.4$ DEEP2 galaxies. Specifically, the ``shock plus precursor''
model, with a shock velocity of $\sim 200$\kms, reproduces the line ratios
quite well. In this case, optical emission lines originate both in hot gas
that has already been shocked, as well as in a pre-shock H~II region that
is being ionized by the radiation field from the cooling shocked gas. Such
a shock ionization model also predicts [SII]$\lambda\lambda
6717+6731$/H$\alpha \sim 0.3$ and [OI]$\lambda 6300$/H$\alpha \sim 0.1$.
We do not obtain significant detections of [SII] or [OI] in any individual
spectra, but find [SII]$\lambda\lambda 6717+6731$/H$\alpha \sim 0.2$ when
we average the spectra of the three DEEP2 $z\sim1.4$ galaxies. With the
current depth of our observations, we cannot rule out the contribution of
shock ionization to the observed emission-line ratios.

Finally, we consider the contribution of an AGN. \citet{ho1993} considered
a class of ``transition objects,'' with emission-line ratios intermediate
between those of star-forming galaxies and AGN. The emission-line ratios
of the $z\sim 1.4$ DEEP2 objects are similar to those of ``transition
objects,'' which are explained by \citet{ho1993} as LINER/Seyfert galaxies
whose integrated spectra are diluted by the contributions from H~II
regions photoionzed by stars. Currently, we have no information about the
X-ray or radio emission properties of our DEEP2 sample, which might help
to identify the presence of an AGN. However, it will be valuable to
measure diagnostic emission-line ratios for DEEP2 galaxies in the Extended
Groth Strip region, where such multi-wavelength data, and optical
morphological information, are available.

We defer a final analysis of the cause for the offset in diagnostic line
ratios of DEEP2 galaxies until a statistical sample is assembled. A larger
high-redshift sample will enable studies of how the diagnostic line ratios
vary as functions of other galaxy properties. Furthermore, more detailed
examination of the small fraction of SDSS galaxies with similar
[OIII]/H$\beta$ and [NII]/H$\alpha$ ratios should aid in interpreting the
DEEP2 sample.  We will compare additional spectral features, stellar
population parameters, and morphological information available for such
SDSS galaxies, relative to those of the SDSS galaxies with more typical
line ratios. At this point, we highlight the implications of the observed
offset in DEEP2 line ratios, relative to the excitation sequence of both
SDSS galaxies and the H~II regions used to formulate the $O3N2$ and $N2$
abundance calibrators.

\subsection{Implications of the Offsets in Emission-line Diagnostics}
\label{sec:bptimp}

Studies of chemical evolution using star-forming galaxies at $z\geq 1$
rely on strong-line abundance indicators. Most of the features of interest
are shifted into the near-IR at these redshifts. It would be most
desirable to measure simultaneously the full set of strong rest-frame
optical emission lines that are available in spectra of nearby
star-forming galaxies \citep{tremonti2004,jansen2000}. However, because of
the limits of current near-IR spectroscopic technology, atmospheric
absorption, and ubiquitous and bright night sky emission lines, only a
subset of H~II region lines can typically be observed simultaneously in
high-redshift galaxies.  In light of these limitations, we have emphasized
the $N2$ and $O3N2$ strong-line abundance indicators for the current work,
especially since these indicators are fairly insensitive to uncertainties
in dust-extinction and flux-calibration. These indicators may represent
the optimal strong-line indicators available for high-redshift galaxies,
using current ground-based instrumentation.

In section~\ref{sec:lzmz}, we emphasized the importance of using the same
empirical abundance indicator (e.g. $R_{23}$, $O3N2$, or $N2$)  for a fair
comparison of galaxy samples at different redshifts. Now, we provide an
important caveat to this argument. When using an empirically-calibrated
abundance indicator such as $O3N2$ or $N2$ for both local and distant
galaxies, it is critical to take into account systematic differences in
the H~II region physical conditions as a function of redshift.
Specifically, the $O3N2$ and $N2$ indicators were calibrated by
\citet{pp2004} using a sample of H~II regions with direct O/H abundance
measurements. These H~II regions apparently follow the same
[OIII]/H$\beta$ vs. [NII]/H$\alpha$ excitation sequence as the
star-forming regions of emission-line galaxies in the SDSS sample of
\citet{tremonti2004}. This similarity suggests that the two samples span
the same range of H~II region physical parameters, in terms of ionizing
spectrum, ionization parameter and metallicity, and that the \cite{pp2004}
calibration should be valid for converting the $O3N2$ values measured from
SDSS spectra into oxygen abundances. Since $z\sim 1.4$ DEEP2 galaxies are
offset from the emission-line sequence of local H~II regions and SDSS
galaxies, systematic biases may result when applying the $O3N2$
calibration of \citet{pp2004} to derive oxygen abundances from their
spectra.

We can further illustrate the problem in two ways. First, we compare $N2$-
and $O3N2$-based abundances for the five galaxies in our DEEP2 sample with
estimates of both quantities. As presented in section~\ref{sec:oh}, the
$N2$-based abundances are all systematically higher than those based on
$O3N2$. For the $z\sim1.4$ sample, the average offset is $0.13$ dex,
whereas for the $z\sim 1.0$ sample, the average offset is $0.08$ dex. This
systematic difference is not hard to understand. Since local H~II regions
form a roughly one-dimensional sequence in the space of [OIII]/H$\beta$
vs. [NII]/H$\alpha$, a given value of $O3N2$ corresponds to a specific
location on the one dimensional sequence, and therefore also determines
the value of $N2$. Furthermore, the $N2$ and $O3N2$ indicators should
yield consistent oxygen abundances for local H~II regions, since both
indicators were calibrated on such a sample. Since the DEEP2 galaxies are
systematically offset from this one-dimensional sequence, their $O3N2$
values correspond to higher values of $N2$, than those of local H~II
regions with similar $O3N2$ values. Therefore, it is the incorrect
assumption of the DEEP2 galaxies following the local H~II region
excitation sequence that leads to the systematic discrepancies between
$N2$- and $O3N2$-based abundances. Second, we use the theoretical models
of \citet{kewley2001} to demonstrate the importance of applying the
correct physical conditions to the interpretation of emission line
spectra. While the PEGASE instantaneous models with $n_e=10\mbox{
cm}^{-3}$ appear to provide the correct input ionizing spectrum to explain
the emission-line sequence of local H~II regions \citep{dopita2000}, the
continuous models with $n_e=350\mbox{ cm}^{-3}$ appear more appropriate
for producing the emission-line ratios observed in nearby starbursts
\citep{kewley2001}. While the absolute abundance calibration of both of
these models may be systematically high in the super-solar regime
\citep{kennicutt2003}, we are simply interested in relative effects here.  
We determine the difference in metallicity for models of fixed ionization
parameter, but different ionizing spectrum, that produce the same emergent
$O3N2$ value. This metallicity difference will indicate the possible
magnitude of error in abundance that results from application of the local
H~II region model to the spectrum of a starburst nucleus -- if indeed the
differences between the two emission-line sequences can be explained by
this difference in ionizing spectrum. To produce the range of $O3N2$
values observed in our sample, the instantaneous H~II region models are
typically $1.5-2.0$ times lower in metallicity than the continuous
starburst models.

Both the systematic differences in $O3N2$- and $N2$-based oxygen
abundances in DEEP2 galaxies, and the error in derived metallicity that
results from using an inappropriate input ionizing spectrum, are within a
factor of 2. These errors in metallicity are smaller than or equal to the
quoted uncertainties in the $N2$ and $O3N2$ abundance indicators, which
result from the intrinsic scatter among the H~II regions that were
included in the calibrations \citep{pp2004}. However, if the relationship
between O/H and $O3N2$ (or $N2$) is systematically offset {\it in the same
sense in all distant star-forming galaxies} from the scatter observed
local H~II regions, this should be taken into account when estimating the
evolution of chemical abundances, the $L-Z$, and $M-Z$ relationships as a
function of redshift.  Our future goal is to determine the cause of the
offset line ratios, and, if possible, to define a local calibration sample
of H~II regions with direct abundance measurements, that follows the same
excitation sequence of the distant starburst galaxies. Finally, galaxies
with high star-formation rates were much more common at $z\geq 1$
\citep{arnouts2004,adelberger2000,erb2003}. Physical differences in H~II
regions with extreme star-formation rates may have a profound impact on
the star-formation process and stellar IMF \citep{stolte2005}. Therefore,
to characterize the build-up of the stellar mass in the universe, we must
have a full understanding of the physical conditions in star-forming
regions in distant galaxies.

\section{SUMMARY}
\label{sec:summary}

We have presented the rest-frame optical spectra of a pilot sample of 12
DEEP2 galaxies at $1.0\leq z \leq 1.5$. These galaxies span almost three
magnitudes in $M_B$ ($-20.3$ to $-23.1$), and are drawn from the set of
galaxies blueward of the observed $(U-B)_0$ color bimodality found in the
DEEP2 survey. They span a range of stellar mass from $4\times 10^9$ to
$2\times 10^{11} M_{\odot}$.  We have measured [NII] and H$\alpha$ fluxes
for the entire sample, and additional [OIII] and H$\beta$ fluxes for a
subset of five objects. Our principal conclusions include the following:

1. The [NII]/H$\alpha$ line ratios for this sample indicate an average gas
phase oxygen abundance of $\sim 0.8 (\rm{O/H}_{\odot})$, with the
application of the $N2$ abundance calibration of \citet{pp2004}. Combining
[NII]/H$\alpha$ and [OIII]/H$\beta$ ratios for the 5 galaxies with all
available measurements, we apply the $O3N2$ abundance calibration of
\citet{pp2004} to obtain an average oxygen abundance of $\sim 0.5
(\rm{O/H}_{\odot})$. For this subsample the $O3N2$ indicator yields
systematically lower oxygen abundances than the $N2$ indicator, by
$0.10-0.15$~dex.

2. Considering the sample of [NII]/H$\alpha$ ratios, absolute $B$
magnitudes, and derived stellar masses, we find evidence that objects with
brighter $M_B$ and larger $M_*$ also have larger [NII]/H$\alpha$ ratios on
average, indicative of higher oxygen abundance. This result suggests that
luminosity-metallicity and mass-metallicity relationships are already in
place among star-forming galaxies at $z\geq 1$, but will require a larger
sample to confirm.

3. A comparison of DEEP2 galaxies with the local SDSS star-forming sample
of \citet{tremonti2004} indicates strong evolution between $z\sim 0$ and
$z\sim 1$ in the relationships among oxygen abundance, rest-frame $B$-band
luminosity, and stellar mass. We highlight results for the subsample of
DEEP2 galaxies with $O3N2$ measurements, characterized by $\langle
12+\log(\mbox{O/H}) \rangle = 8.4$, $\langle M_B \rangle=-21.9$, and
$\langle M_*\rangle = 3 \times 10^{10}M_{\odot}$. Applying the $O3N2$
abundance indicator consistently to both samples, we find that SDSS
galaxies with similar O/H to the DEEP2 sample average are $\sim 5.2$
magnitudes fainter and $\sim 40$ times less massive. Conversely, SDSS
galaxies with similar luminosities to the average of the DEEP2 sample have
O/H abundances that are 0.35~dex higher. SDSS galaxies with similar masses
to the DEEP2 average are characterized by O/H abundances 0.30~dex higher.

4. The clustering strengths and number densities of blue DEEP2 galaxies
with strong emission lines \citep{coil2004,faber2005} are consistent with
a scenario in which the majoriy of our DEEP2 sample will remain blue and
star-forming by $z\sim 0$. In this scenario, the descendants of our DEEP2
sample will be included in the sample of \citet{tremonti2004}. To explain
how the $z\geq 1$ sample evolves to lie on the locally-observed
luminosity-metallicity and mass-metallicity relationships, we use the
observed evolution in $M_{B*}$ in the rest-frame $B$-band luminosity
function between $z\sim 1$ and $z\sim 0$ \citep{faber2005}. This evolution
consists of 1.3 magnitudes of fading for the DEEP2 galaxies, to an average
magnitude of $\langle M_B \rangle = -20.6$.  SDSS galaxies with $M_B \sim
-20.6$ have typical stellar masses of $4\times 10^{10} M_{\odot}$, a
factor of two larger than in our DEEP2 sample, and stellar $M/L_B$ ratios
$6-7$ times higher. To lie on the SDSS $L-Z$ relationship, the DEEP2
oxygen abundances must increase on average by 0.35 dex. The combined
increase in stellar mass and metallicity is roughly consistent with the
expectations of a closed-box chemical evolution model, which is not very
surprising, since the masses of the DEEP2 galaxies are larger than the
mass at which feedback appears to be important
\citep{garnett2002,kauffmann2003,tremonti2004}. Though they have similar
[NII]/H$\alpha$ ratios and rest-frame $B$ luminosities, the sample of
$z\sim 2$ star-forming galaxies with $K_s\leq 20.0$ presented by
\citet{shapley2004} will follow a very different evolutionary path from
that of objects in our DEEP2 sample. These $z\sim 2$ galaxies have stellar
masses and $M/L_B$ ratios that are already $\sim 3$ times higher on
average than the DEEP2 galaxies presented here, and are significantly more
strongly clustered than blue DEEP2 galaxies in general. They will no
longer be forming stars by $z\sim 1$, and their descendants will not be
included in the sample of \citet{tremonti2004}.

5. We compute the standard emission-line diagnostic, [OIII]/H$\beta$ vs.
[NII]/H$\alpha$, for the five DEEP2 galaxies for which such measurements
are available. Bright DEEP2 galaxies at $z\sim 1.4$ are significantly
offset from the tight excitation sequence followed by SDSS star-forming
galaxies and local H~II regions. A similar offset is observed in both
nearby starburst galaxies \citep{kewley2001} as well as star-forming
galaxies at $z\sim 2$ (Erb et al., in preparation). These galaxies all
feature elevated star-formation rates, relative to typical values in the
SDSS sample. In particular, the DEEP2 $z\sim 1.4$ galaxies have H$\alpha$
star-formation rates of $10-15 M_{\odot}\mbox{yr}^{-1}$. We consider
several causes for the observed offset in line ratios, including
differences in the ionizing spectrum, ionization parameter, and electron
density; and the effects of shocks and AGN. Differences in the H~II region
physical conditions must be taken into account when applying
empirically-calibrated abundance indicators, such as $N2$ and $O3N2$, to
galaxy samples at different redshifts.

6. Our pilot sample of DEEP2 galaxies demonstrates the feasibility of
chemical abundance studies at $1.0 \leq z \leq 1.5$. We used long-slit
spectroscopy on a ground-based 10-meter class telescope, with relatively
short integration times in single near-IR bandpasses.  New infrared
spectrographs are now available on large telescopes featuring broader
simultaneous wavelength coverage (e.g., the Gemini Near Infrared
Spectrograph), and the availability of multi-object near-infrared
spectrographs is imminent. The DEEP2 redshift survey, with its extensive
set of precise nebular [OII] redshifts, provides an ideal sample with
which to explore the metal content and H~II region physics in distant
star-forming galaxies using these new ground-based technologies.

\smallskip
We are indebted to the DEEP2 team, whose significant efforts in
establishing such a tremendous spectroscopic sample at $z\sim 1$ made this
project possible. We also thank the staff at the W.~M. Keck Observatory
for their assistance with the NIRSPEC observations, in particular Jim
Lyke, whose extensive help was invaluable. We also thank Christy Tremonti,
Lisa Kewley, and Max Pettini
for enlightening conversations, with special thanks to
Christy for providing us with a table of SDSS
line fluxes, absolute magnitudes and stellar masses, allowing for a
straightforward comparison with local galaxies.  We thank Chris Conselice
and Richard Ellis, who carried out the near-infrared photometric
observations that enabled modeling of stellar masses. AES acknowledges
support from the Miller Foundation for Basic Research in Science. We wish
to extend special thanks to those of Hawaiian ancestry on whose sacred
mountain we are privileged to be guests. Without their generous
hospitality, most of the observations presented herein would not have been
possible.


\clearpage
\begin{deluxetable}{llllcccccc}
\tablewidth{0pt}
\tabletypesize{\footnotesize}
\tablecaption{Galaxies Observed With Keck~II/NIRSPEC\label{tab:obs}}
\tablehead{
\colhead{Galaxy} &
\colhead{R.A. (J2000)} &
\colhead{Dec. (J2000)} &
\colhead{$z_{H\alpha}$} &
\colhead{$R$} &
\colhead{$B-R$} &
\colhead{$R-I$} &
\colhead{$K_s$} &
\colhead{Exposure (s)} &
\colhead{Band}
}
\startdata
32038085 & 23:29:57.68 &  00:22:28.66 &  1.0183 & 23.33 &  0.57 &  0.88 & 19.10   & 2$\times$900 & J \\
 & & & & & & & & 2$\times$900 & N1 \\
32038132 & 23:29:58.74 &  00:22:23.88 &  1.0190 & 23.87 &  0.44 &  0.62 & 20.17   & 2$\times$900 & J \\
 & & & & & & & & 2$\times$900 & N1 \\
32020728 & 23:30:08.53 &  00:10:09.60 &  1.0446 & 23.43 &  0.11 &  0.49 & 20.36   & 3$\times$900 & J \\
 & & & & & & & & 3$\times$900 & N1 \\
42009739 & 02:29:35.76 &  00:25:07.25 &  1.3067 & 23.12 &  0.18 &  0.33 & \nodata & 3$\times$900 & H \\
42008627 & 02:29:54.34 &  00:23:56.42 &  1.3625 & 22.32 &  0.31 &  0.31 & 19.01   & 1$\times$900 & H \\
 & & & & & & & & 2$\times$900 &  J \\
42008700 & 02:29:53.06 &  00:23:48.46 &  1.3890 & 23.95 &  0.24 &  0.43 & 20.25   & 3$\times$900 & H \\
 & & & & & & & & 2$\times$900 &  J \\
32100778 & 23:30:38.74 &  00:14:07.57 &  1.3925 & 22.58 &  0.58 &  0.61 & 18.31   & 2$\times$900 & H \\
32025514 & 23:30:38.28 &  00:13:55.90 &  1.3938 & 23.89 &  0.43 &  0.53 & 18.66   & 6$\times$900 & H \\
32037003 & 23:30:13.73 &  00:20:05.22 &  1.3983 & 22.40 &  0.65 &  0.49 & \nodata & 2$\times$900 & H \\
42055947 & 02:29:23.25 &  00:45:27.88 &  1.3987 & 21.90 &  0.35 &  0.29 & 18.83   & 3$\times$900 & H \\
 & & & & & & & & 2$\times$900 &  J \\
32036994 & 23:30:13.68 &  00:20:24.42 &  1.3990 & 23.87 &  0.30 &  0.33 & \nodata & 2$\times$900 & H \\
42008219 & 02:30:04.48 &  00:25:00.73 &  1.4010 & 23.58 &  0.15 &  0.40 & 21.41   & 3$\times$900 & H \\
 & & & & & & & & 3$\times$900 &  J \\
\enddata
\end{deluxetable}

\clearpage
\begin{landscape}
\begin{deluxetable}{llrrrrcccccc}
\tablewidth{0pt}
\small
\tabletypesize{\footnotesize}
\tablecaption{Emission lines and Physical Quantities\label{tab:meas}}
\tablehead{
\colhead{Galaxy} &
\colhead{$z_{H\alpha}$} &
\colhead{$F_{\rm{H}\beta}$\tablenotemark{a}} &
\colhead{$F_{\rm{[OIII]} \lambda 5007}$\tablenotemark{a}} &
\colhead{$F_{\rm{H}\alpha}$\tablenotemark{a}} &
\colhead{$F_{\rm{[NII]} \lambda 6584}$\tablenotemark{a}} &
\colhead{$12+\log (\rm{O/H})_{(N2)}$\tablenotemark{b}} &
\colhead{$12+\log (\rm{O/H})_{(O3N2)}$\tablenotemark{c}} &
\colhead{$L_{\rm{H}\alpha}$\tablenotemark{d}} &
\colhead{$SFR_{\rm{H}\alpha}$\tablenotemark{e}} &
\colhead{$M_B$} &
\colhead{$\log(M_*/M_{\odot})$\tablenotemark{f}}
}
\startdata
32038085 &  1.0183 & $1.9\pm0.3$ & $2.9\pm0.3$  & $14.4\pm0.4$ &  $3.5\pm0.3$ & $8.55\pm0.18$ & $8.48\pm0.14$ & 0.8 &  6 & -21.28 & $10.67\pm0.07$ \\
32038132 &  1.0190 & \nodata & \nodata  &  $5.6\pm0.2$ &  $1.1\pm0.2$ & $8.51\pm0.19$ & \nodata       & 0.3 &  2 & -20.38 & $10.08\pm0.20$ \\
32020728 &  1.0446 & $4.0\pm0.2$ & $12.4\pm0.2$ & $10.7\pm0.1$ &  $1.1\pm0.2$ & $8.33\pm0.18$ & $8.25\pm0.14$ & 0.6 &  5 & -20.72 &  $9.60\pm0.16$ \\
42009739 &  1.3067 &  \nodata    &  \nodata     & $12.4\pm0.2$ &  $1.4\pm0.2$ & $8.37\pm0.18$ &  \nodata      & 1.3 & 10 & -21.67 &  \nodata       \\
42008627 &  1.3625 & $3.8\pm0.4$ & $7.7\pm0.4$  & $18.5\pm0.7$ &  $6.0\pm0.7$ & $8.62\pm0.18$ & $8.48\pm0.14$ & 2.1 & 16 & -22.59 & $10.51\pm0.02$ \\
42008700 &  1.3890 & \nodata & \nodata  &  $7.3\pm0.2$ &  $2.1\pm0.2$ & $8.59\pm0.18$ &  \nodata      & 0.9 &  7 & -21.27 & $10.40\pm0.18$ \\
32100778 &  1.3925 &  \nodata    &  \nodata     &  $6.5\pm0.3$ &  $2.8\pm0.3$ & $8.70\pm0.18$ &  \nodata      & 0.8 &  6 & -23.05 & $11.21\pm0.13$ \\
32025514 &  1.3938 &  \nodata    &  \nodata     &  $6.5\pm0.3$ &  $2.1\pm0.3$ & $8.62\pm0.18$ &  \nodata      & 0.8 &  6 & -21.57 & $11.34\pm0.05$ \\
32037003 &  1.3983 &  \nodata    &  \nodata     & $17.6\pm0.4$ &  $4.0\pm0.4$ & $8.53\pm0.18$ &  \nodata      & 2.1 & 17 & -22.99 &  \nodata       \\
42055947 &  1.3987 & $8.6\pm0.6$ & $18.2\pm0.6$ & $17.3\pm0.2$ &  $3.7\pm0.3$ & $8.52\pm0.18$ & $8.41\pm0.14$ & 2.1 & 16 & -23.05 & $10.61\pm0.04$ \\
32036994 &  1.3990 &  \nodata    &  \nodata     & $15.6\pm0.3$ &  $3.5\pm0.3$ & $8.53\pm0.18$ &  \nodata      & 1.9 & 15 & -21.17 &  \nodata       \\
42008219 &  1.4010 & $3.8\pm0.4$ & $8.4\pm0.3$  & $11.2\pm0.2$ &  $3.3\pm0.2$ & $8.60\pm0.18$ & $8.45\pm0.14$ & 1.3 & 11 & -21.62 &  $9.95\pm0.16$ \\
\enddata
\tablenotetext{a}{$\:$Line flux and random error in units of
$10^{-17}\mbox{ erg s}^{-1}\mbox{ cm}^{-2}$.}
\tablenotetext{b}{Oxygen abundance deduced from the $N2$
relationship presented in \citet{pp2004}.}
\tablenotetext{c}{Oxygen abundance deduced from the $O3N2$
relationship presented in \citet{pp2004}.}
\tablenotetext{d}{$\Ha$ luminosity in units of $10^{42}\mbox{ erg s}^{-1}$.}
\tablenotetext{e}{Star-formation rate in units of
$M_{\odot}\mbox{yr}^{-1}$, calculated from $L_{\Ha}$ using the conversion
of \citet{kennicutt1998}.}
\tablenotetext{f}{Stellar mass and uncertainty estimated using the methods described
in section~\ref{sec:lzmzobs}.}
\end{deluxetable}
\clearpage
\end{landscape}

\end{document}